\documentclass[12pt]{article}
\usepackage[latin9]{inputenc}
\usepackage[a4paper]{geometry}
\usepackage{float}
\usepackage{booktabs}
\usepackage{bm}
\usepackage{amsmath}
\usepackage{amsthm}
\usepackage{amssymb}
\usepackage{setspace}
\usepackage[authoryear]{natbib}
\makeatletter

\providecommand{\tabularnewline}{\\}

\theoremstyle{plain}
\newtheorem{prop}{\protect\propositionname}
\theoremstyle{plain}
\newtheorem{cor}{\protect\corollaryname}

\renewcommand{\hat}{\widehat}
\newtheorem{assumption}{Assumption}

\date{August 10, 2022}

\providecommand{\propositionname}{Proposition}

\makeatother

\providecommand{\corollaryname}{Corollary}
\providecommand{\propositionname}{Proposition}

\begin{document}
\title{Selecting Valid Instrumental Variables in Linear Models with Multiple
Exposure Variables: Adaptive Lasso and the Median-of-Medians Estimator }
\author{Xiaoran Liang$^{a}$, Eleanor Sanderson$^{b,d}$, Frank Windmeijer$^{b,c}$
\\
 $^{a}${\small{}School of Economics, University of Bristol, UK}\\
 $^{b}${\small{}MRC Integrative Epidemiology Unit, University of
Bristol, UK}\\
 $^{c}${\small{}Dept of Statistics and Nuffield College, University
of Oxford, UK}\\
 $^{d}${\small{}Population Health Science Institute, Bristol Medical
School, University of Bristol, UK}}
\maketitle
\begin{abstract}
\noindent \baselineskip=15pt In a linear instrumental variables (IV)
setting for estimating the causal effects of multiple confounded exposure/treatment
variables on an outcome, we investigate the adaptive Lasso method
for selecting valid instrumental variables from a set of available
instruments that may contain invalid ones. An instrument is invalid
if it fails the exclusion conditions and enters the model as an explanatory
variable. We extend the results as developed in \citet{WindmeijeretalJASA2019}
for the single exposure model to the multiple exposures case. In particular
we propose a median-of-medians estimator and show that the conditions
on the minimum number of valid instruments under which this estimator
is consistent for the causal effects are only moderately stronger
than the simple majority rule that applies to the median estimator
for the single exposure case. The adaptive Lasso method which uses
the initial median-of-medians estimator for the penalty weights achieves
consistent selection with oracle properties of the resulting IV estimator.
This is confirmed by some Monte Carlo simulation results. We apply
the method to estimate the causal effects of educational attainment
and cognitive ability on body mass index (BMI) in a Mendelian Randomization
setting. 
\end{abstract}
\noindent \textbf{\small{}Keywords:}{\small{} Causal inference; Adaptive
Lasso; Instrumental variables; Invalid instruments; Mendelian randomization;
Median-of-medians estimator }{\small\par}

\thispagestyle{empty}

\baselineskip=20pt

\pagebreak{}

\pagenumbering{arabic} \setcounter{page}{1}

\section{Introduction}

Instrumental variable (IV) methods are widely used to determine the
causal effect of a treatment/exposure on an outcome when their relationship
is potentially confounded by unobserved factors. In IV estimation,
it is crucial that instruments are valid. This requires that (a) the
instruments must be associated with the exposure variable (the relevance
condition), and (b) the only pathway from the instruments to the outcome
is through the exposure; the instruments do not have direct effects
on the outcome nor affect the outcome through unobservables (the exclusion
conditions). In our setting, we are concerned with the situation where
we have a fixed, but large number of available instruments that satisfy
the relevance condition. However, some of the instruments may violate
the exclusion conditions and hence are invalid. If we include these
invalid instruments in IV estimation, the resulting estimator will
be inconsistent. It is therefore important to have selection methods
that consistently selects the valid instruments.

Previous work has addressed the IV selection problem in the case of
a single exposure variable. \citet{KangetalJASA2016} establish the
model setup for this IV selection. They develop the identification
conditions and propose a selection method based on the Lasso \citep{tibshirani1996regression}.
\citet{WindmeijeretalJASA2019} propose a method based on the adaptive
Lasso \citep{Zou2006adaptive} under the assumption that more than
half of the candidate instruments are valid; the so-called majority
rule. The median of the instrument-specific estimates is then a consistent
estimator of the causal effect and can be used for the penalization
of the adaptive Lasso, resulting in consistent selection of the valid
instruments and oracle properties of the post-selection IV estimator,
meaning that the IV estimator behaves in large samples as if the set
of valid instruments were known \citep{fan2001variable}. \citet{GuoetalJRSSB2018}
refine the identification condition proposed by \citet{KangetalJASA2016}
and establish the sufficient and necessary identification condition
which is the plurality rule. It states that the valid instruments
form the largest group, where instruments form a group if the instrument-specific
estimators for the causal effect converge to the same value, and is
hence a relaxation of the majority rule. The Hard Thresholding with
Voting method proposed by \citet{GuoetalJRSSB2018} can achieve consistent
selection under the plurality rule. Also assuming the plurality rule,
\citet{windmeijer2021confidence} propose the Confidence Interval
method which result in consistent selection, and has as an advantage
over the Hard Thresholding with Voting method that the number of instruments
selected as valid in the Confidence Interval method decreases monotonically
when decreasing the tuning parameter.

Unlike the existing literature above, we consider here the case of
multiple, potentially confounded exposure variables. This setting
can be motivated by recent Mendelian Randomization (MR) studies in
epidemiology. In MR studies, genetic variants are used as instruments
for estimating the causal effect of a modifiable exposure on a health-related
outcome. In many cases, there are additional exposure variables that
need to be considered apart from the primary exposure. For example,
\citet{SandersonetalIJE2019} estimate the effect of educational attainment
on body mass index (BMI) conditional on cognitive ability. Both educational
attainment and cognitive ability are confounded by unobserved factors
that affect both the outcome and the exposure variables. Therefore,
a method to select the valid instruments needs to take account of
the multiple exposure variables problem.

We contribute to the literature by extending the adaptive Lasso method
in \citet{WindmeijeretalJASA2019} to allow for multiple exposure
variables. The main issue for the adaptive Lasso is to have an initial
consistent estimator of the causal effects that can be used for the
penalization. For the single exposure case, the median of the instrument-specific
estimates of the causal effect is a consistent estimator when more
than 50\% of the instruments are valid and satisfies the conditions
for oracle properties when used in the adaptive Lasso for instrument
selection. This could simply be extended for the multiple exposure
case to the medians of all just-identified estimates of the causal
effects. A just-identified estimator is one where the number of instruments
used is equal to the number of exposure variables. Let $k_{x}$ and
$k_{z}$ denote the number of exposure and instrumental variables
respectively. Then there are $p=\binom{k_{z}}{k_{x}}$ just-identified
estimators of the causal effects and if more than $50\%$ of these
are consistent, then the medians of these $p$ estimators are consistent.
Let $k_{\mathcal{V}}$ denote the number of valid instruments. Under
a strong relevance assumption that each set of just-identifying instruments
are jointly relevant for all exposure variables, this majority rule
then implies that $\binom{k_{\mathcal{V}}}{k_{x}}>$$\frac{1}{2}$$\binom{k_{z}}{k_{x}}$.
As an example, with $k_{x}=2$ and $k_{z}=21$, we have $210$ just-identifying
pairs of instruments, of which more than $105$ need to be pairs of
valid instruments. This implies that for this naive median estimator
at least $16$ instruments need to be valid.

We propose a novel median-of-medians estimator which we show to be
a consistent estimator of the causal effects and which utilizes the
available information better, in the sense that it requires less instruments
to be valid for consistency compared to the naive median estimator.
We show in Section \ref{sec:MM} that for $k_{x}\geq2$, the median-of-medians
estimator is consistent if $k_{\mathcal{V}}>\frac{k_{z}+k_{x}-1}{2}$.
This condition is a (weakly) weaker condition on the number of valid
instruments than for the naive median estimator, with the difference
increasing in $k_{z}$. For the case of $k_{x}=2$ this results in
the condition that $k_{\mathcal{V}}>\frac{k_{z}+1}{2}$, and so for
$k_{z}=21$ this implies that at least $12$ instruments need to be
valid. In other words, whereas the naive median estimator allows in
this case for a maximum of $5$ instruments to be invalid, the median-of-medians
estimator is still consistent with $9$ invalid instruments. The condition
for the median-of-medians estimator is a natural progression of the
condition for the single-exposure majority rule that $k_{\mathcal{V}}>\frac{k_{z}}{2}$,
and for $k_{x}=2$, it is only stricter when $k_{z}$ is odd. For
$k_{x}>2$, the estimator is strictly speaking a median-of-medians-of-medians....
estimator, but for brevity we will call it the median-of-medians estimator
for all $k_{x}$.

The assumption that all just-identifying sets of instruments are jointly
relevant for all exposure variables may not hold in practice. In our
application, genetic variants that are candidate instruments for educational
attainment and cognitive ability are identified in separate genome-wide
association studies (GWAS) and there is very little overlap of genetic
variants between the two traits. We can adjust the median-of-medians
estimator for this block structure of the instruments by only considering
in this case the just-identifying pairs of instruments, where each
pair contains one instrument from each group.

The paper is structured as follows. Section \ref{sec:model} introduces
the model, IV estimation and the adaptive Lasso IV selection method
for selecting the valid/invalid instruments. Section \ref{sec:MM}
introduces the median-of-medians estimator and derives its properties.
In Section \ref{sec:downward} we discuss the median-of-medians estimator
based consistent selection and oracle properties of the adaptive Lasso
method, also combining it with the downward testing procedure for
model selection proposed by \citet{Andrews1999}. In Section \ref{sec: extension},
we introduce the block structure variation of the method that accounts
for violation of the full rank assumption. Section \ref{sec: simulation}
presents some Monte Carlo simulation results. In Section \ref{sec: application},
we apply our method to Mendelian randomization and estimate the causal
effects of educational attainment and cognitive ability on BMI. Section
\ref{sec: conclusion} concludes.

\textbf{Notation.} In the remainder of the paper, let $\left\Vert \left\{ .\right\} \right\Vert _{q}$
denote the $l_{q}$-norm of a vector. For a matrix $\mathbf{X}_{n\times k_{x}}$
with full column rank, let $\mathbf{P}_{X}=\mathbf{X}(\mathbf{X}^{T}\mathbf{X})^{-1}\mathbf{X}^{T}$
and $\mathbf{M}_{X}=\mathbf{I}_{n}-\mathbf{P}_{X}$, where $\mathbf{I}_{n}$
is the $n$-dimensional identity matrix. For a general matrix $\mathbf{A}$,
$r\left(\mathbf{A}\right)$ denotes its rank. Convergence in probability
and distribution are indicated by $\stackrel{p}{\rightarrow}$ and
$\stackrel{d}{\rightarrow}$ respectively.

\section{Model, IV Estimation and Adaptive Lasso}

\label{sec:model}

We have an i.i.d. sample $\{Y_{i},\mathbf{X}_{i}^{T},\mathbf{Z}_{i}^{T}\}_{i=1}^{n}$,
where $Y_{i}$ is the outcome of interest for observation $i$, $\mathbf{X}_{i}$
is a $k_{x}$-vector of exposure variables, $\mathbf{Z}_{i}$ is a
$k_{z}$-vector of putative instrumental variables and $n$ is the
sample size. As in \citet{GuoetalJRSSB2018}, \citet{WindmeijeretalJASA2019}
and \citet{windmeijer2021confidence}, we follow \citet{KangetalJASA2016}
who, starting from the additive linear constant effects model of \citet{Holland1988},
arrived at the observed data model for the random sample given by
\begin{equation}
Y_{i}=\mathbf{X}_{i}^{T}\bm{\beta}+\mathbf{Z}_{i}^{T}\bm{\alpha}+U_{i},\label{outcome equation}
\end{equation}
where $\bm{\beta}$ is the causal parameter vector of interest, and
with $\mathbb{E}[U_{i}|\mathbf{Z}_{i}]=0$, but $\mathbf{X}_{i}$
may be confounded by $U_{i}$. The parameter vector $\bm{\alpha}$
captures the violations of the exclusion restriction. Formally, following
the definition of invalid instruments as in \citet[p 797]{GuoetalJRSSB2018},
for $j\in1,...,k_{z}$, an instrument $Z_{j}$ is invalid if $\alpha_{j}\neq0$
and valid if $\alpha_{j}=0$. Let $\mathcal{V}$ and $\mathcal{A}$
be the sets of indices of the valid and invalid instruments respectively:
$\mathcal{V}=\left\{ j:\alpha_{j}=0\right\} $, $\mathcal{A}=\left\{ j:\alpha_{j}\neq0\right\} $,
with dimensions $k_{\mathcal{V}}$ and $k_{\mathcal{A}}$ respectively,
then $k_{z}=k_{\mathcal{V}}+k_{\mathcal{A}}$.

Let $\mathbf{y}$ be the $n$-vector of $n$ observations on $\left\{ Y_{i}\right\} $,
and let $\mathbf{X}$ and $\mathbf{Z}$ be the $n\times k_{x}$ and
$n\times k_{z}$ matrices of the exposure variables and candidate
instrumental variables, respectively. Let $\mathbf{Z}_{\mathcal{V}}$
and $\mathbf{Z}_{\mathcal{A}}$ denote the $n\times k_{\mathcal{V}}$
and $n\times k_{\mathcal{A}}$ matrices of valid and invalid instruments.
The oracle model is the model where the set of invalid instruments
is known, and is hence given by 
\[
\mathbf{y}=\mathbf{X}\bm{\beta}+\mathbf{Z}_{\mathcal{A}}\bm{\alpha}_{\mathcal{A}}+\mathbf{u},
\]
where $\mathbf{u}$ is the $n$-vector with elements $\left\{ U_{i}\right\} $.
The so-called first-stage regression model of $\mathbf{X}$ on $\mathbf{Z}$
is given by 
\begin{align*}
\mathbf{X} & =\mathbf{Z}\bm{\Pi}+\mathbf{E}\\
 & =\mathbf{Z}_{\mathcal{V}}\bm{\Pi}_{\mathcal{V}}+\mathbf{Z}_{\mathcal{A}}\bm{\Pi}_{\mathcal{A}}+\mathbf{E},
\end{align*}
where $\boldsymbol{\Pi}=\left[\boldsymbol{\Pi}_{\mathcal{V}}\,\,\boldsymbol{\Pi}_{\mathcal{A}}\right]=\left(\mathbb{E}\left[\mathbf{Z}_{i}\mathbf{Z}_{i}^{T}\right]\right)^{-1}\mathbb{E}\left[\mathbf{Z}_{i}\mathbf{X}_{i}^{T}\right]$
and $\mathbb{E}\left[\mathbf{E}_{i}|\mathbf{Z}_{i}\right]=0$.

We assume the instrument relevance condition for the oracle model
holds:

\begin{assumption} \label{rel} Relevance: $r$$\left(\boldsymbol{\Pi}_{\mathcal{V}}\right)=k_{x}$.
\end{assumption}

We further make the standard assumptions as in \citet{WindmeijeretalJASA2019}:

\begin{assumption} \label{assexp} $\mathbb{E}\left[\mathbf{Z}_{i}\mathbf{Z}_{i}^{T}\right]=\mathbf{Q}_{zz}$,
with $\mathbf{Q}_{zz}$ a finite and full rank matrix; $\mathbb{E}\left[\mathbf{Z}_{i.}\mathbf{X}_{i.}^{T}\right]=\mathbf{Q}_{zx}$,
with $\mathbf{Q}_{zx}$ a finite matrix. \end{assumption}

\begin{assumption} \label{clt} $\frac{1}{\sqrt{n}}\boldsymbol{Z}^{T}\boldsymbol{u}\overset{d}{\rightarrow}N\left(0,\bm{\Sigma}_{zu}\right)$
as $n\rightarrow\infty$ , with $\boldsymbol{\Sigma}_{zu}$ a finite
and full rank matrix.\end{assumption}

\begin{assumption} \label{homo} Homoskedasticity: $\mathbb{E}\left[U_{i}^{2}|\mathbf{Z}_{i}\right]=\sigma_{u}^{2}$.\end{assumption}

\noindent It follows that under under the homoskedasticity assumption,
$\bm{\Sigma}_{zu}=\sigma_{u}^{2}\boldsymbol{Q}_{zz}$.

\subsection{IV Estimation}

\label{subsec:IV-Estimation}

Let $\boldsymbol{\theta}^{or}=\left(\boldsymbol{\beta}^{T}\,\,\boldsymbol{\alpha}_{\mathcal{A}}^{T}\right)^{T}$
and $\mathbf{R}=\left[\mathbf{X}\,\,\mathbf{Z}_{\mathcal{A}}\right]$.
A standard two-stage least squares (2sls) IV estimator of $\boldsymbol{\theta}^{or}$
is defined as 
\[
\hat{\boldsymbol{\theta}}_{2sls}^{or}=\arg\min_{\boldsymbol{\theta}}\left(\mathbf{y}-\mathbf{R}\bm{\theta}\right)^{T}\mathbf{Z}\left(\mathbf{Z}^{T}\mathbf{Z}\right)^{-1}\mathbf{Z}^{T}\left(\mathbf{y}-\mathbf{R}\bm{\theta}\right),
\]
 resulting in 
\begin{align*}
\hat{\boldsymbol{\theta}}_{2sls}^{or} & =\left(\mathbf{R}^{T}\mathbf{P}_{Z}\mathbf{R}\right)^{-1}\mathbf{R}^{T}\mathbf{P}_{Z}\mathbf{y}\\
 & =\left(\hat{\mathbf{R}}^{T}\hat{\mathbf{R}}\right)^{-1}\hat{\mathbf{R}}^{T}\mathbf{y},
\end{align*}
where $\hat{\mathbf{R}}=\left[\hat{\mathbf{X}}\,\,\mathbf{Z}_{\mathcal{A}}\right]$,
with $\hat{\mathbf{X}}=\mathbf{Z}\hat{\boldsymbol{\Pi}}$, and $\hat{\boldsymbol{\Pi}}=\left(\mathbf{Z}^{T}\mathbf{Z}\right)^{-1}\mathbf{Z}^{T}\mathbf{X}$.
Under Assumptions \ref{rel}-\ref{homo} the 2sls estimator is asymptotically
efficient, and its limiting distribution is given by 
\begin{equation}
\sqrt{n}\left(\hat{\boldsymbol{\theta}}_{2sls}^{or}-\boldsymbol{\theta}\right)\stackrel{d}{\rightarrow}N\left(0,\sigma_{u}^{2}\left(\boldsymbol{Q}_{zx}^{T}\boldsymbol{Q}_{zz}^{-1}\boldsymbol{Q}_{zx}\right)^{-1}\right).\label{eq:thetaorlim}
\end{equation}

From standard partitioned regression results, we can express the 2sls
estimators for $\boldsymbol{\beta}$ and $\boldsymbol{\alpha}_{\mathcal{A}}$
as 
\begin{align}
\hat{\boldsymbol{\beta}}_{2sls}^{or} & =\left(\hat{\mathbf{X}}^{T}\mathbf{M}_{Z_{\mathcal{A}}}\hat{\mathbf{X}}\right)^{-1}\hat{\mathbf{X}}^{T}\mathbf{M}_{Z_{\mathcal{A}}}\mathbf{y},\label{eq:betaor}\\
\hat{\boldsymbol{\alpha}}_{\mathcal{A}}^{or} & =\left(\mathbf{Z}_{\mathcal{A}}^{T}\mathbf{M}_{\hat{X}}\mathbf{Z}_{\mathcal{A}}\right)^{-1}\mathbf{Z}_{\mathcal{A}}^{T}\mathbf{M}_{\hat{X}}\mathbf{y}.\label{eq:alphaor}
\end{align}

When $k_{\mathcal{V}}>k_{x}$, the test for overidentifying restrictions
is a test for $H_{0}:\mathbb{E}\left[\mathbf{Z}_{i}U_{i}\right]=0$.
The \citet{Sargan1958} test statistic is given by 
\begin{equation}
S\left(\hat{\boldsymbol{\theta}}_{2sls}^{or}\right)=\frac{\left(\mathbf{y}-\mathbf{R}\hat{\boldsymbol{\theta}}_{2sls}^{or}\right)^{T}\mathbf{Z}\left(\mathbf{Z}^{T}\mathbf{Z}\right)^{-1}\mathbf{Z}^{T}\left(\mathbf{y}-\mathbf{R}\hat{\boldsymbol{\theta}}_{2sls}^{or}\right)}{\left(\mathbf{y}-\mathbf{R}\hat{\boldsymbol{\theta}}_{2sls}^{or}\right)^{T}\left(\mathbf{y}-\mathbf{R}\hat{\boldsymbol{\theta}}_{2sls}^{or}\right)/n},\label{eq:SarganS}
\end{equation}
which, under the null, converges in distribution to a $\chi_{k_{\mathcal{V}}-k_{x}}^{2}$
distributed random variable under Assumptions \ref{rel}-\ref{homo}.

Let a selection of $k_{\mathcal{A}^{*}}$ instruments classified as
invalid be denoted $\mathbf{\ensuremath{Z}_{\mathcal{A}^{*}}}$, with
$k_{z}-k_{\mathcal{A}^{*}}\geq k_{x}$. The corresponding model is
given by 
\begin{align*}
\mathbf{y} & =\mathbf{X}\boldsymbol{\beta}+\mathbf{Z}_{\mathcal{A}^{*}}\boldsymbol{\alpha}_{\mathcal{A}^{*}}+\mathbf{u}^{*}\\
 & =\mathbf{R}^{*}\boldsymbol{\theta}^{*}+\mathbf{u}^{*}.
\end{align*}
Then, if the set contains all invalid instruments, such that $\mathbf{Z}_{\mathcal{A}}\mathbf{\subseteq Z_{\mathcal{A}^{*}}}$,
it follows that the IV estimator for $\boldsymbol{\beta}$ is consistent
and normal and, for $k_{z}-k_{\mathcal{A}^{*}}>k_{x}$, $S\left(\hat{\boldsymbol{\theta}}_{2sls}^{*}\right)\stackrel{d}{\rightarrow}\chi_{k_{z}-k_{x}-k_{\mathcal{A}^{*}}}^{2}$.
Alternatively, under the plurality rule that the valid instruments
form the largest group, it follows that for all sets with $k_{\mathcal{A}^{*}}=k_{\mathcal{A}}$,
if $\mathbf{Z}_{\mathcal{A}^{*}}\neq\mathbf{Z}_{\mathcal{A}}$, then
the IV estimator for $\boldsymbol{\beta}$ is inconsistent and $S\left(\hat{\boldsymbol{\theta}}_{2sls}^{*}\right)=O_{p}\left(n\right)$.

\subsection{Adaptive Lasso}

\label{sec: method}

Based on the definition of a valid instrument, selection of the valid
instruments is equivalent to identifying which entries in $\bm{\alpha}$
are zero. For this purpose, we consider using the adaptive Lasso to
estimate $\bm{\alpha}$, as the Lasso will shrink some entries in
$\bm{\alpha}$ to exactly zero. Hence, we can obtain estimators for
$\mathcal{V}$ and $\mathcal{A}$ from the adaptive Lasso estimator
for $\bm{\alpha}$, which we denote by $\hat{\bm{\alpha}}_{ad}$.
The estimators for $\mathcal{V}$ and $\mathcal{A}$ are then $\hat{\mathcal{V}}=\{j:\hat{\alpha}_{ad,j}=0\}$
and $\hat{\mathcal{A}}=\{j:\hat{\alpha}_{ad,j}\neq0\}$.

\citet{KangetalJASA2016} introduced the Lasso method for IV selection
for the single exposure case. \citet{WindmeijeretalJASA2019} showed
that the Lasso irrepresentable condition (see \citealp{Zhao2006model},
and \citealp{Zou2006adaptive}) could be violated, depending on the
relative strengths of the invalid and valid instruments, leading to
inconsistent selection of the valid/invalid instruments. They adopted
the adaptive Lasso estimator of \citet{Zou2006adaptive}. Let now
$\boldsymbol{\theta}=\left(\boldsymbol{\beta}^{T}\,\,\boldsymbol{\alpha}^{T}\right)^{T}$,
then the penalized objective function is based on the 2sls criterion
and the adaptive Lasso estimator is given by 
\begin{equation}
\hat{\bm{\theta}}_{ad}=\underset{\bm{\beta},\bm{\alpha}}{\mathrm{\arg\min}}\frac{1}{2}\left\Vert \left\{ \mathbf{P}_{Z}(\mathbf{y}-\mathbf{X}\bm{\beta}-\mathbf{Z}\bm{\alpha})\right\} \right\Vert _{2}^{2}+\lambda_{n}\sum_{j=1}^{k_{z}}\frac{\left|\alpha_{j}\right|}{\left|\hat{\alpha}_{j}\right|^{\nu}},\label{ass:ch2orad}
\end{equation}
where $\hat{\boldsymbol{\alpha}}$, with $j$-th element equal to
$\hat{\alpha}_{j}$, is an initial estimator of $\boldsymbol{\alpha}$,
and $\nu>0$. As $\bm{\beta}$ is not penalized, the adaptive Lasso
estimator for \textbf{$\boldsymbol{\alpha}$} can be obtained as 
\begin{equation}
\hat{\bm{\alpha}}_{ad}=\underset{\bm{\alpha}}{\mathrm{\arg\min}}\frac{1}{2}\left\Vert {\mathbf{y}-\widetilde{\mathbf{Z}}\bm{\alpha}}\right\Vert _{2}^{2}+\lambda_{n}\sum_{j=1}^{k_{z}}\frac{\left|\alpha_{j}\right|}{\left|\hat{\alpha}_{j}\right|^{\nu}},\label{alasso selection}
\end{equation}
where $\widetilde{\mathbf{Z}}=\mathbf{M}_{\hat{X}}\mathbf{Z}$, see
\citet{KangetalJASA2016} and \citet{WindmeijeretalJASA2019}.

$\lambda_{n}$ is the tuning parameter controlling the strength of
the penalization. A larger $\lambda_{n}$ leads to more entries in
$\bm{\alpha}$ being shrunk to zero, which implies that the adaptive
Lasso selects more instruments as valid. From Theorem 2 and Remark
1 in \citet{Zou2006adaptive} the adaptive Lasso estimator for $\bm{\alpha}$,
as defined in (\ref{alasso selection}) has oracle properties and
hence selects the valid instruments consistently under the following
assumptions: 

\begin{assumption}\label{ass:alphahat}$\hat{\boldsymbol{\alpha}}\stackrel{p}{\rightarrow}\boldsymbol{\alpha}$
and $\sqrt{n}\left(\hat{\boldsymbol{\alpha}}-\boldsymbol{\alpha}\right)=O_{p}\left(1\right)$.
\end{assumption}

\begin{assumption}\label{ass:lambdan} $\lambda_{n}=o(\sqrt{n})$,
$n^{\frac{\nu-1}{2}}\lambda_{n}\rightarrow\infty$. \end{assumption}

The intuition for $\hat{\alpha}_{j}$ is clear. As a consistent estimator
for $\alpha_{j}$, $\hat{\alpha}_{j}$ will be close to zero when
$\alpha_{j}=0$. Since $\hat{\alpha}_{j}$ enters in the denominator
in (\ref{alasso selection}), a value close to zero will produce a
large penalty weight, and make it more likely that $\hat{\alpha}_{ad,j}$
is equal to zero. The oracle properties then apply to $\hat{\bm{\theta}}_{ad}$
and hence the estimator of interest $\hat{\boldsymbol{\beta}}_{ad}$.

Clearly, a crucial component for the application of the adaptive Lasso
estimator is the initial consistent estimator of $\boldsymbol{\alpha}$.
We propose an initial consistent estimator of $\boldsymbol{\beta}$,
the median-of-medians estimator as described in the next section,
from which the required estimator for $\boldsymbol{\alpha}$ can be
derived.

\section{The Median-of-Medians Estimator}

\label{sec:MM}

For the single exposure case, \citet{WindmeijeretalJASA2019}, following
\citet{Han2008Detecting}, showed that the median of the instrument
specific, just-identified estimators for $\beta$ is a consistent
estimator of $\beta$ when more than 50\% of the instruments are valid,
i.e.\ $k_{\mathcal{V}}>\frac{k_{z}}{2}$. These just-identified estimators
are the IV, or 2sls estimators in the model specifications 
\[
\mathbf{y}=\mathbf{x}\beta_{j}+\mathbf{Z}_{\left\{ -j\right\} }\bm{\alpha}_{\left\{ -j\right\} }+\mathbf{u}_{j},
\]
for $j=1,\ldots,k_{z}$, and where $\mathbf{Z}_{\left\{ -j\right\} }=\mathbf{Z}\setminus\left\{ \mathbf{z}_{j}\right\} $
is the full set of instruments with the $j$-th instrument omitted,
which is used as the excluded instrument for $\mathbf{x}$, and $\mathbf{u}_{j}=\mathbf{z}_{j}\alpha_{j}+\mathbf{u}$,
see also \citet{windmeijer2021confidence}. Let $\hat{\beta}_{j}$
denote this IV estimator for $\beta$ that treats the $j$-th instrument
as the valid instrument and all other instruments as invalid. Provided
all instruments are relevant, $\pi_{j}\neq0$ for $j=1,\ldots,k_{z}$
in $\mathbf{x}=\mathbf{Z}\boldsymbol{\pi}+\mathbf{e}$, it follows
that $\mathbf{z}_{j}\alpha_{j}=\left(\mathbf{x}-\mathbf{Z}_{\left\{ -j\right\} }\boldsymbol{\pi}_{\left\{ -j\right\} }-\mathbf{e}\right)\frac{\alpha_{j}}{\pi_{j}}$.
Then for the valid instruments, $j\in\mathcal{V}$, $\hat{\beta}_{j}$
is a consistent and normal estimator of $\beta$, whereas for the
invalid instruments, $j\in\mathcal{A}$, $\hat{\beta}_{j}$ is a consistent
and normal estimator of $\beta+\frac{\alpha_{j}}{\pi_{j}}$, and hence
an inconsistent estimator of $\beta$. It then follows that the median
estimator, given by
\begin{equation}
\hat{\beta}_{m}=\text{median}\left\{ \hat{\beta}_{j}\right\} _{j=1}^{k_{z}}\label{eq:bm}
\end{equation}
is a consistent estimator of $\beta$ if $k_{\mathcal{V}}>\frac{k_{z}}{2}$
and \citet{WindmeijeretalJASA2019} show that then $\sqrt{n}\left(\hat{\beta}_{m}-\beta\right)=O_{p}\left(1\right)$.
A consistent estimator for $\boldsymbol{\alpha}$ is then given by
\[
\hat{\boldsymbol{\alpha}}_{m}=\left(\mathbf{Z}{}^{T}\mathbf{Z}\right)^{-1}\mathbf{Z}^{T}\left(\mathbf{y}-\mathbf{x}\hat{\beta}_{m}\right),
\]
with also $\sqrt{n}\left(\hat{\boldsymbol{\alpha}}_{m}-\boldsymbol{\alpha}\right)=O_{p}\left(1\right)$,
satisfying the conditions for oracle properties of the adaptive Lasso
estimator, leading to consistent selection of the valid and invalid
instruments and oracle properties of $\hat{\beta}_{ad}$.

We can extend the median estimator to the case where there are multiple
exposure variables, $k_{x}\geq2$. We initially assume that all $p=\binom{k_{z}}{k_{x}}$
just-identifying sets of instruments are jointly relevant for all
exposure variables. Denote the just-identifying sets of instruments
by $\mathbf{Z}_{s}$, for $s=1,\ldots,p$. The just-identified model
specifications are then given by
\begin{align}
\mathbf{y} & =\mathbf{X}\boldsymbol{\beta}_{s}+\mathbf{Z}_{\left\{ -s\right\} }\bm{\alpha}_{\left\{ -s\right\} }+\mathbf{u}_{s}\label{eq:jimod}\\
\mathbf{X} & =\mathbf{Z}_{s}\bm{\Pi}_{s}+\mathbf{Z}_{\left\{ -s\right\} }\bm{\Pi}_{\left\{ -s\right\} }+\mathbf{E},\label{eq:fss}
\end{align}
where $\mathbf{Z}_{\left\{ -s\right\} }=\mathbf{Z}\setminus\left\{ \mathbf{Z}_{s}\right\} $
and $\mathbf{u}_{s}=\mathbf{Z}_{s}\boldsymbol{\alpha}_{s}+\mathbf{u}$.
These relevance conditions can then be stated as follows,

\begin{assumption}\label{ass:justrank} Relevance of just-identifying
sets. Let the $p=\binom{k_{z}}{k_{x}}$ sets of just-identifying instruments
be denoted $\left\{ \mathbf{Z}_{s}\right\} _{s=1}^{p}$ and let $\bm{\Pi}_{s}$
be defined as in (\ref{eq:fss}). Then all these sets are jointly
relevant for all exposure variables, $r\left(\bm{\Pi}_{s}\right)=k_{x}$,
for $s=1,\ldots,p$. \end{assumption}

It then follows that $\mathbf{Z}_{s}\boldsymbol{\alpha}_{s}=\left(\mathbf{X}-\mathbf{Z}_{\left\{ -s\right\} }\bm{\Pi}_{\left\{ -s\right\} }-\mathbf{E}\right)\boldsymbol{\Pi}_{s}^{-1}\boldsymbol{\alpha}_{s}$
and so the estimands for the just-identified IV estimators are given
by $\boldsymbol{\beta}_{s}=\text{\ensuremath{\boldsymbol{\beta}+\boldsymbol{\Pi}_{s}^{-1}\boldsymbol{\alpha}_{s}}}$,
resulting in consistent and normal IV estimators for the sets that
contain valid instruments only, with $\boldsymbol{\alpha}_{s}=\mathbf{0}$,
and inconsistent estimators of $\boldsymbol{\beta}$ for all other
sets. Let $\hat{\boldsymbol{\beta}}_{m}$ denote the medians of the
$p$ just-identified estimators, then it follows directly from the
results in \citet{WindmeijeretalJASA2019} that $\hat{\boldsymbol{\beta}}_{m}$
is a consistent estimator of $\boldsymbol{\beta}$ and $\sqrt{n}\left(\hat{\boldsymbol{\beta}}_{m}-\boldsymbol{\beta}\right)=O_{p}\left(1\right)$
if more than half of all just-identified sets are sets of valid instruments
only, or $\binom{k_{\mathcal{V}}}{k_{x}}>\frac{p}{2}$, see also \citet{Apfel2019}.

\subsection{$k_{x}=2$}

In comparison to this naive median estimator, we can allow for a weaker
condition on the number of valid instruments required to obtain an
initial consistent estimator based on the just-identified estimators.
To this end we propose the median-of-medians estimator. For ease of
exposition, we first consider the $k_{x}=2$ case. Consider for each
instrument $j=1,\ldots,k_{z}$, all just-identifying sets of instruments
that contain instrument $j$. There are $k_{z}-1$ such sets for each
$j$. Denote the just-identified estimators based on the sets that
contain $j$ by $\hat{\boldsymbol{\beta}}_{\ell}^{j}$, $\ell=1,\ldots,k_{z},\ell\neq j$.
If instrument $j$ is invalid, $j\in\mathcal{A}$, then none of the
$\hat{\boldsymbol{\beta}}_{\ell}^{j}$ are consistent estimators of
$\boldsymbol{\beta}$. If instrument $j$ is valid, $j\in\mathcal{V}$,
then the $k_{\mathcal{V}}-1$ sets of instruments containing $j$
and another valid instrument result in consistent and normal IV estimators
of $\boldsymbol{\beta}$, whereas the remaining $k_{z}-k_{\mathcal{V}}$
sets contain invalid instruments resulting in inconsistent IV estimators.
Let 
\begin{equation}
\hat{\boldsymbol{\beta}}_{m}^{j}=\text{median}\left\{ \hat{\boldsymbol{\beta}}_{\ell}^{j}\right\} _{\ell=1,\ell\neq j}^{k_{z}},\label{eq:bmj}
\end{equation}
where the medians are taking element wise, so $\hat{\beta}_{m,q}^{j}=\text{median}\left\{ \hat{\beta}_{\ell,q}^{j}\right\} _{\ell=1,\ell\neq j}^{k_{z}}$
for $q=1,2$.

For $\hat{\boldsymbol{\beta}}_{m}^{j}$ to be a consistent estimator
of $\boldsymbol{\beta}$ for a valid instrument $j\in\mathcal{V}$,
we need the following further assumption,

\begin{assumption}\label{ass:jmajority} Condition on number of valid
instruments. For $k_{x}=2$, the number of valid instruments $k_{\mathcal{V}}$
satisfies $\left(k_{\mathcal{V}}-1\right)>\frac{k_{z}-1}{2}$, or
equivalently $k_{\mathcal{V}}>\frac{k_{z}+1}{2}$.\end{assumption}

Under Assumption \ref{ass:jmajority}, it follows that for a valid
instrument $j\in\mathcal{V}$ the majority rule is satisfied. This
implies that more than half of the just-identifying sets of instruments
containing $j$ are sets of valid instruments only, and hence the
result follows straightforwardly that $\hat{\boldsymbol{\beta}}_{m}^{j}$
is a consistent estimator of $\boldsymbol{\beta}$ and $\sqrt{n}\left(\hat{\boldsymbol{\beta}}_{m}^{j}-\boldsymbol{\beta}\right)=O_{p}\left(1\right)$,
from the results and proof of Theorem 1 in \citet{WindmeijeretalJASA2019}.
The result is stated in the following proposition, with the proof
given in the Appendix. 
\begin{prop}
\label{prop:bmj}For $k_{x}=2$, let for each instrument $j=1,\ldots,k_{z}$,
the median estimator $\hat{\boldsymbol{\beta}}_{m}^{j}$ be defined
as in (\ref{eq:bmj}). Then, under Assumptions \ref{assexp}, \ref{clt},
\ref{ass:justrank} and \ref{ass:jmajority}, for valid instruments
$j\in\mathcal{V}$, $\hat{\boldsymbol{\beta}}_{m}^{j}$ is a consistent
estimator of $\boldsymbol{\beta}$, $\hat{\boldsymbol{\beta}}_{m}^{j}\stackrel{p}{\rightarrow}\boldsymbol{\beta}$,
and $\sqrt{n}\left(\hat{\boldsymbol{\beta}}_{m}^{j}-\boldsymbol{\beta}\right)=O_{p}\left(1\right)$. 
\end{prop}
Under the conditions of Proposition \ref{prop:bmj} it follows that
$k_{\mathcal{V}}$ out of $k_{z}$ estimators $\left\{ \hat{\boldsymbol{\beta}}_{m}^{j}\right\} _{j=1}^{k_{z}}$
are consistent estimators of $\boldsymbol{\beta}$. It then follows
that the median-of-medians estimator, defined as 
\begin{equation}
\hat{\boldsymbol{\beta}}_{mm}=\text{median}\left\{ \hat{\boldsymbol{\beta}}_{m}^{j}\right\} _{j=1}^{k_{z}},\label{eq:bmm}
\end{equation}
where the medians are again taken element wise, is a consistent estimator
of $\boldsymbol{\beta}$ and $\sqrt{n}\left(\hat{\boldsymbol{\beta}}_{mm}-\boldsymbol{\beta}\right)=O_{p}\left(1\right)$
if $k_{\mathcal{V}}>\frac{k_{z}}{2}$, but this condition is implied
by Assumption \ref{ass:jmajority}. We formally state this result
in the following proposition, with the proof presented in the Appendix. 
\begin{prop}
\label{prop:mm} For $k_{x}=2$, let the median-of-medians estimator
$\hat{\boldsymbol{\beta}}_{mm}$ be defined as in (\ref{eq:bmm}),
then under the conditions of Proposition \ref{prop:bmj} it follows
that $\hat{\boldsymbol{\beta}}_{mm}\stackrel{p}{\rightarrow}\boldsymbol{\beta}$,
and $\sqrt{n}\left(\hat{\boldsymbol{\beta}}_{mm}-\boldsymbol{\beta}\right)=O_{p}\left(1\right)$. 
\end{prop}
As an illustration of the median-of-medians estimator, we consider
the case with $k_{x}=2$, $k_{z}=7$ and $k_{\mathcal{V}}=5>\frac{k_{z}+1}{2}=4$,
and so Assumption \ref{ass:jmajority} is satisfied. Let instruments
$1$ and $2$ be the invalid ones, so $\mathcal{A}=\left\{ 1,2\right\} $
and $\mathcal{V}=\left\{ 3,4,5,6,7\right\} $. Table \ref{tab: mmillus}
lists the just-identified estimators for $\beta_{q}$, $q=1,2$, and
they are estimated using each IV pair. The valid instruments and consistent
estimators are indicated in boldface. 
\begin{table}[h]
\begin{centering}
\caption{\label{tab: mmillus} Illustration of the median-of-medians estimator
of $\beta_{q}$, $q=1,2$.}
\par\end{centering}
\medskip{}

\begin{centering}
\begin{tabular}{ccccccccc}
\toprule 
Instruments  & \multicolumn{1}{c}{$1$} & $2$  & $\boldsymbol{3}$  & $\boldsymbol{4}$  & $\boldsymbol{5}$  & $\boldsymbol{6}$  & $\boldsymbol{7}$  & \tabularnewline
\midrule 
$1$  & \multicolumn{1}{c}{} & $\hat{\beta}_{1,q}^{2}$  & $\hat{\beta}_{1,q}^{3}$  & $\hat{\beta}_{1,q}^{4}$  & $\hat{\beta}_{1,q}^{5}$  & $\hat{\beta}_{1,q}^{6}$  & $\hat{\beta}_{1,q}^{7}$  & \tabularnewline
$2$  & $\hat{\beta}_{2,q}^{1}$  &  & $\hat{\beta}_{2,q}^{3}$  & $\hat{\beta}_{2,q}^{4}$  & $\hat{\beta}_{2,q}^{5}$  & $\hat{\beta}_{2,q}^{6}$  & $\hat{\beta}_{2,q}^{7}$  & \tabularnewline
$\boldsymbol{3}$  & $\hat{\beta}_{3,q}^{1}$  & $\hat{\beta}_{3,q}^{2}$  &  & $\boldsymbol{\hat{\beta}_{3,q}^{4}}$  & $\boldsymbol{\hat{\beta}_{3,q}^{5}}$  & $\boldsymbol{\hat{\beta}_{3,q}^{6}}$  & $\boldsymbol{\hat{\beta}_{3,q}^{7}}$  & \tabularnewline
$\boldsymbol{4}$  & $\hat{\beta}_{4,q}^{1}$  & $\hat{\beta}_{4,q}^{2}$  & $\boldsymbol{\hat{\beta}_{4,q}^{3}}$  &  & $\boldsymbol{\hat{\beta}_{4,q}^{5}}$  & $\boldsymbol{\hat{\beta}_{4,q}^{6}}$  & $\boldsymbol{\hat{\beta}_{4,q}^{7}}$  & \tabularnewline
$\boldsymbol{5}$  & $\hat{\beta}_{5,q}^{1}$  & $\hat{\beta}_{5,q}^{2}$  & $\boldsymbol{\hat{\beta}_{5,q}^{3}}$  & $\boldsymbol{\hat{\beta}_{5,q}^{4}}$  &  & $\boldsymbol{\hat{\beta}_{5,q}^{6}}$  & $\boldsymbol{\hat{\beta}_{5,q}^{7}}$  & \tabularnewline
$\boldsymbol{6}$  & $\hat{\beta}_{6,q}^{1}$  & $\hat{\beta}_{6,q}^{2}$  & $\boldsymbol{\hat{\beta}_{6,q}^{3}}$  & $\boldsymbol{\hat{\beta}_{6,q}^{4}}$  & $\boldsymbol{\hat{\beta}_{6,q}^{5}}$  &  & $\boldsymbol{\hat{\beta}_{6,q}^{7}}$ & \tabularnewline
$\boldsymbol{7}$  & $\hat{\beta}_{7,q}^{1}$  & $\hat{\beta}_{7,q}^{2}$  & $\boldsymbol{\hat{\beta}_{7,q}^{3}}$  & $\boldsymbol{\hat{\beta}_{7,q}^{4}}$  & $\boldsymbol{\hat{\beta}_{7,q}^{5}}$  & $\boldsymbol{\hat{\beta}_{7,q}^{6}}$ &  & \tabularnewline
\midrule 
median  & $\hat{\beta}_{m,q}^{1}$  & $\hat{\beta}_{m,q}^{2}$  & $\boldsymbol{\hat{\beta}_{m,q}^{3}}$  & $\boldsymbol{\hat{\beta}_{m,q}^{4}}$  & $\boldsymbol{\hat{\beta}_{m,q}^{5}}$  & $\boldsymbol{\hat{\beta}_{m,q}^{6}}$  & $\boldsymbol{\hat{\beta}_{m,q}^{7}}$  & $\boldsymbol{\hat{\beta}_{mm,q}}$\tabularnewline
\bottomrule
\end{tabular}
\par\end{centering}
\begin{centering}
{\small{}Notes: $k_{x}=2$, $k_{z}=7$, $\mathcal{V}=\left\{ 3,4,5,6,7\right\} $,
$\mathcal{A}=\left\{ 1,2\right\} $. Valid instruments and consistent
estimators are displayed in boldface. }{\small\par}
\par\end{centering}
 
\end{table}

For the general case, the just-identified estimator $\hat{\boldsymbol{\beta}}_{\ell}^{j}$
is a consistent estimator of $\boldsymbol{\beta}$ if and only if
both instruments $j$ and $\ell$ are valid. Hence, all the estimators
of $\beta_{q}$, $q=1,2$, in the columns for instruments $1$ and
$2$ are inconsistent as at least one of the invalid instruments is
involved in the estimation and the resulting median estimators $\hat{\beta}_{m,q}^{1}$
and $\hat{\beta}_{m,q}^{2}$ are inconsistent. For instruments $3$-$7$
more than half of the $k_{z}-1$ estimators in each column are consistent
as here we have $k_{\mathcal{V}}-1>\frac{k_{z}-1}{2}$. Hence, the
median estimators $\hat{\beta}_{m,q}^{3},\ldots,\hat{\beta}_{m,q}^{7}$
are all consistent. Now, we take the median of all these column median
estimators (as shown in the last row of Table~\ref{tab: mmillus}),
i.e.\ $\hat{\beta}_{mm,q}=\text{median}\left(\hat{\beta}_{m,q}^{1},...,\hat{\beta}_{m,q}^{7}\right)$.
The assumption $k_{\mathcal{V}}>\frac{k_{z}+1}{2}$ implies $k_{\mathcal{V}}>\frac{k_{z}}{2}$.
Thus, more than half of the column median estimators $\hat{\beta}_{m,q}^{1},...,\hat{\beta}_{m,q}^{7}$
are consistent and therefore the median of these median estimators
$\hat{\beta}_{mm,q}$ is also consistent. Therefore, for $k_{x}=2$,
under the assumption $k_{\mathcal{V}}>\frac{k_{z}+1}{2}$ the median-of-medians
estimator is consistent even if we have no knowledge about which of
the instruments are valid. 

For comparison, for the naive median estimator to be consistent, the
condition $\binom{k_{\mathcal{V}}}{k_{x}}>\frac{1}{2}\binom{k_{z}}{k_{x}}$
implies here that $k_{\mathcal{V}}>4$, so only one instrument is
allowed to be invalid. Increasing $k_{z}$ to $k_{z}=100$, the condition
for the median-of-medians estimator is that $k_{\mathcal{V}}>50.5$,
whereas for the naive median estimator this is $k_{\mathcal{V}}>70$.
For $k_{x}=2$, Assumption \ref{ass:jmajority} for the median-of-medians
estimator that $k_{\mathcal{V}}>\frac{k_{z}+1}{2}$ is only stronger
for the minimum number of valid instruments required than the simple
majority rule for the single exposure model when $k_{z}$ is odd,
with the difference then equal to $1$.

\subsection{General $k_{x}$ }

We can extend the results for the median-of-medians estimator for
the $k_{x}=2$ case to the $k_{x}>2$ case, where the estimator becomes
a median-of-medians-of-medians-of.... estimator, but we simply refer
to it as the median-of-medians estimator for brevity.

Let the instrument set be denoted $S=\left\{ 1,\ldots,k_{z}\right\} $.
Consider the $l_{1}=\prod_{j=0}^{k_{x}-2}\left(k_{z}-j\right)$ sets
of $k_{x}-1$ instruments $\mathcal{L}_{1}=\left\{ L_{1,r_{1}}\right\} _{r_{1}=1}^{l_{1}}$.
This is a tree structure, with the $k_{z}$ instruments at the top
of the tree, and all interactions branching down the tree. For example,
for $k_{x}=3$ the sets $\left\{ 1,2\right\} $ and $\left\{ 2,1\right\} $
are both included in $\mathcal{L}_{1}$. See Table \ref{tab:Illusm2}
for an illustation for the $k_{x}=3$ case. For each of the sets $L_{1,r_{1}}$
and for each instrument $\ell\in S\setminus L_{1,r_{1}}$, let $\hat{\boldsymbol{\beta}}_{\ell}^{L_{1,r_{1}}}$
denote the just-identified IV estimator of $\boldsymbol{\beta}$,
using the $k_{x}$ instruments $\left\{ L_{1,r_{1}},\ell\right\} $.
For brevity, for a general set $A\supseteq S$ denote $S_{A}:=S\setminus A$.
For each set $L_{1,r_{1}}$, we have thus $k_{z}-k_{x}+1$ just identified
estimators. If the set $L_{1,r_{1}}$ is a set of $k_{x}-1$ valid
instruments then the majority of $\left\{ \hat{\boldsymbol{\beta}}_{\ell}^{L_{1,r_{1}}}\right\} _{\ell\in S_{L_{1,r_{1}}}}$
are consistent and normal if there are additionally more than $\frac{k_{z}-k_{x}+1}{2}$
valid instruments. Therefore, if $k_{\mathcal{V}}>\frac{k_{z}+k_{x}-1}{2}$
and $L_{1,r_{1}}$ is a set of valid instruments, then 
\[
\hat{\boldsymbol{\beta}}_{m}^{L_{1,r_{1}}}=\text{median}\left\{ \hat{\boldsymbol{\beta}}_{\ell}^{L_{1,r_{1}}}\right\} _{\ell\in S_{L_{1,r_{1}}}}
\]
is a consistent estimator of $\boldsymbol{\beta}$, and from Proposition
\ref{prop:bmj} it follows that $\hat{\boldsymbol{\beta}}_{m}^{L_{1,r_{1}}}\stackrel{p}{\rightarrow}\boldsymbol{\beta}$
and $\sqrt{\ensuremath{n}}\left(\hat{\boldsymbol{\beta}}_{m}^{L_{1,r_{1}}}-\boldsymbol{\beta}\right)=O_{p}\left(1\right)$.
If $k_{x}=1$ then $\mathcal{L}_{1}=\emptyset$ and the standard median
estimator is obtained.

\begin{table}
\caption{\label{tab:Illusm2} Illustration of median-of-medians-of-medians
estimator of $\beta_{q}$, $q=1,2,3$}

\begin{centering}
\begin{tabular}{cccc|ccc|c|ccc|ccc}
\hline 
$\mathcal{L}_{2}$ &  &  & \multicolumn{1}{c}{} &  &  & \multicolumn{1}{c}{} & \multicolumn{1}{c}{\textbf{1}} &  &  & \multicolumn{1}{c}{} &  &  & \tabularnewline
\hline 
$\mathcal{L}_{1}$ &  & \textbf{12} &  &  & \textbf{13} &  & $\boldsymbol{\hat{\beta}_{mm,q}^{1}}$ &  & \textbf{14} &  &  & \textbf{1}5 & \tabularnewline
\hline 
 &  & $\boldsymbol{\hat{\beta}_{m,q}^{12}}$ &  &  & $\boldsymbol{\hat{\beta}_{m,q}^{13}}$ &  &  &  & $\boldsymbol{\hat{\beta}_{m,q}^{14}}$ &  &  & $\hat{\beta}_{m,q}^{15}$ & \tabularnewline
 & \textbf{123} & \textbf{124} & \textbf{12}5 & \textbf{132} & \textbf{134} & \textbf{13}5 &  & \textbf{142} & \textbf{143} & \textbf{14}5 & \textbf{1}5\textbf{2} & \textbf{1}5\textbf{3} & \textbf{1}5\textbf{4}\tabularnewline
 & $\boldsymbol{\hat{\beta}_{3,q}^{12}}$ & $\boldsymbol{\hat{\beta}_{4,q}^{12}}$ & $\hat{\beta}_{5,q}^{12}$ & $\boldsymbol{\hat{\beta}_{2,q}^{13}}$ & $\boldsymbol{\hat{\beta}_{4,q}^{13}}$ & $\hat{\beta}_{5,q}^{13}$ &  & $\boldsymbol{\hat{\beta}_{2,q}^{14}}$ & $\boldsymbol{\hat{\beta}_{3,q}^{14}}$ & $\hat{\beta}_{5,q}^{14}$ & $\hat{\beta}_{2,q}^{15}$ & $\hat{\beta}_{3,q}^{15}$ & $\hat{\beta}_{4,q}^{15}$\tabularnewline
\hline 
$\mathcal{L}_{2}$ &  &  & \multicolumn{1}{c}{} &  &  & \multicolumn{1}{c}{} & \multicolumn{1}{c}{\textbf{2}} &  &  & \multicolumn{1}{c}{} &  &  & \tabularnewline
\hline 
$\mathcal{L}_{1}$ &  & \textbf{21} &  &  & \textbf{23} &  & $\boldsymbol{\hat{\beta}_{mm,q}^{2}}$ &  & \textbf{24} &  &  & \textbf{2}5 & \tabularnewline
\hline 
 &  & $\boldsymbol{\hat{\beta}_{m,q}^{21}}$ &  &  & $\boldsymbol{\hat{\beta}_{m,q}^{23}}$ &  &  &  & $\boldsymbol{\hat{\beta}_{m,q}^{24}}$ &  &  & $\hat{\beta}_{m,q}^{25}$ & \tabularnewline
 & \textbf{213} & \textbf{214} & \textbf{21}5 & \textbf{231} & \textbf{234} & \textbf{23}5 &  & \textbf{241} & \textbf{243} & \textbf{24}5 & \textbf{2}5\textbf{1} & \textbf{2}5\textbf{3} & \textbf{2}5\textbf{4}\tabularnewline
 & $\boldsymbol{\hat{\beta}_{3,q}^{21}}$ & $\boldsymbol{\hat{\beta}_{4,q}^{21}}$ & $\hat{\beta}_{5,q}^{21}$ & $\boldsymbol{\hat{\beta}_{1,q}^{23}}$ & $\boldsymbol{\hat{\beta}_{4,q}^{23}}$ & $\hat{\beta}_{5,q}^{23}$ &  & $\boldsymbol{\hat{\beta}_{1,q}^{24}}$ & $\boldsymbol{\hat{\beta}_{3,q}^{24}}$ & $\hat{\beta}_{5,q}^{24}$ & $\hat{\beta}_{1,q}^{25}$ & $\hat{\beta}_{3,q}^{25}$ & $\hat{\beta}_{4,q}^{25}$\tabularnewline
\hline 
$\mathcal{L}_{2}$ &  &  & \multicolumn{1}{c}{} &  &  & \multicolumn{1}{c}{} & \multicolumn{1}{c}{\textbf{3}} &  &  & \multicolumn{1}{c}{} &  &  & \tabularnewline
\hline 
$\mathcal{L}_{1}$ &  & \textbf{31} &  &  & \textbf{32} &  & $\boldsymbol{\hat{\beta}_{mm,q}^{3}}$ &  & \textbf{34} &  &  & \textbf{3}5 & \tabularnewline
\hline 
 &  & $\boldsymbol{\hat{\beta}_{m,q}^{31}}$ &  &  & $\boldsymbol{\hat{\beta}_{m,q}^{32}}$ &  &  &  & $\boldsymbol{\hat{\beta}_{m,q}^{34}}$ &  &  & $\hat{\beta}_{m,q}^{35}$ & \tabularnewline
 & \textbf{312} & \textbf{314} & \textbf{31}5 & \textbf{321} & \textbf{324} & \textbf{32}5 &  & \textbf{341} & \textbf{342} & \textbf{34}5 & \textbf{3}5\textbf{1} & \textbf{3}5\textbf{2} & \textbf{3}5\textbf{4}\tabularnewline
 & $\boldsymbol{\hat{\beta}_{2,q}^{31}}$ & $\boldsymbol{\hat{\beta}_{4,q}^{31}}$ & $\hat{\beta}_{5,q}^{31}$ & $\boldsymbol{\hat{\beta}_{1,q}^{32}}$ & $\boldsymbol{\hat{\beta}_{4,q}^{32}}$ & $\hat{\beta}_{5,q}^{32}$ &  & $\boldsymbol{\hat{\beta}_{1,q}^{34}}$ & $\boldsymbol{\hat{\beta}_{2,q}^{34}}$ & $\hat{\beta}_{5,q}^{34}$ & $\hat{\beta}_{1,q}^{35}$ & $\hat{\beta}_{2,q}^{35}$ & $\hat{\beta}_{4,q}^{35}$\tabularnewline
\hline 
$\mathcal{L}_{2}$ &  &  & \multicolumn{1}{c}{} &  &  & \multicolumn{1}{c}{} & \multicolumn{1}{c}{\textbf{4}} &  &  & \multicolumn{1}{c}{} &  &  & \tabularnewline
\hline 
$\mathcal{L}_{1}$ &  & \textbf{41} &  &  & \textbf{42} &  & $\boldsymbol{\hat{\beta}_{mm,q}^{4}}$ &  & \textbf{43} &  &  & \textbf{4}5 & \tabularnewline
\hline 
 &  & $\boldsymbol{\hat{\beta}_{m,q}^{41}}$ &  &  & $\boldsymbol{\hat{\beta}_{m,q}^{42}}$ &  &  &  & $\boldsymbol{\hat{\beta}_{m,q}^{43}}$ &  &  & $\hat{\beta}_{m,q}^{45}$ & \tabularnewline
 & \textbf{412} & \textbf{413} & \textbf{41}5 & \textbf{421} & \textbf{423} & \textbf{42}5 &  & \textbf{431} & \textbf{432} & \textbf{43}5 & \textbf{4}5\textbf{1} & \textbf{4}5\textbf{2} & \textbf{4}5\textbf{3}\tabularnewline
 & $\boldsymbol{\hat{\beta}_{2,q}^{41}}$ & $\boldsymbol{\hat{\beta}_{3,q}^{41}}$ & $\hat{\beta}_{5,q}^{41}$ & $\boldsymbol{\hat{\beta}_{1,q}^{42}}$ & $\boldsymbol{\hat{\beta}_{3,q}^{42}}$ & $\hat{\beta}_{5,q}^{42}$ &  & $\boldsymbol{\hat{\beta}_{1,q}^{43}}$ & $\boldsymbol{\hat{\beta}_{2,q}^{43}}$ & $\hat{\beta}_{5,q}^{43}$ & $\hat{\beta}_{1,q}^{45}$ & $\hat{\beta}_{2,q}^{45}$ & $\hat{\beta}_{3,q}^{45}$\tabularnewline
\hline 
$\mathcal{L}_{2}$ &  &  & \multicolumn{1}{c}{} &  &  & \multicolumn{1}{c}{} & \multicolumn{1}{c}{5} &  &  & \multicolumn{1}{c}{} &  &  & \tabularnewline
\hline 
$\mathcal{L}_{1}$ &  & 5\textbf{1} &  &  & 5\textbf{2} &  & $\hat{\beta}_{mm,q}^{5}$ &  & 5\textbf{3} &  &  & 5\textbf{4} & \tabularnewline
\hline 
 &  & $\hat{\beta}_{m,q}^{51}$ &  &  & $\hat{\beta}_{m,q}^{52}$ &  &  &  & $\hat{\beta}_{m,q}^{53}$ &  &  & $\hat{\beta}_{m,q}^{54}$ & \tabularnewline
 & 5\textbf{12} & 5\textbf{13} & 5\textbf{14} & 5\textbf{21} & 5\textbf{23} & 5\textbf{24} &  & 5\textbf{31} & 5\textbf{32} & 5\textbf{34} & 5\textbf{41} & 5\textbf{42} & 5\textbf{43}\tabularnewline
 & $\hat{\beta}_{2,q}^{51}$ & $\hat{\beta}_{3,q}^{51}$ & $\hat{\beta}_{4,q}^{51}$ & $\hat{\beta}_{1,q}^{52}$ & $\hat{\beta}_{3,q}^{52}$ & $\hat{\beta}_{4,q}^{52}$ &  & $\hat{\beta}_{1,q}^{53}$ & $\hat{\beta}_{2,q}^{53}$ & $\hat{\beta}_{4,q}^{53}$ & $\hat{\beta}_{1,q}^{54}$ & $\hat{\beta}_{2,q}^{54}$ & $\hat{\beta}_{3,q}^{54}$\tabularnewline
\hline 
 &  &  & \multicolumn{1}{c}{} &  &  & \multicolumn{1}{c}{} & \multicolumn{1}{c}{$\boldsymbol{\hat{\beta}_{mm_{2},q}}$} &  &  & \multicolumn{1}{c}{} &  &  & \tabularnewline
\hline 
\end{tabular}
\par\end{centering}
{\small{}Notes: $k_{x}=3$, $k_{z}=5$, $\mathcal{V}=\left\{ 1,2,3,4\right\} $,
$\mathcal{A}=\left\{ 5\right\} $. $k_{\mathcal{V}}=4>\frac{k_{z}+k_{x}-1}{2}=3.5$.
Valid instruments and consistent estimators are displayed in boldface.
$\mathcal{L}_{a}$ are the }$l_{a}=\prod_{j=0}^{k_{x}-a-1}\left(k_{z}-j\right)${\small{}
sets of $k_{x}-a$ instruments for $a=1,2$. As examples, $\boldsymbol{\hat{\beta}_{3,q}^{12}}$
is the just-identified IV estimator using the instruments $\left\{ 1,2,3\right\} $,
$\boldsymbol{\hat{\beta}_{m,q}^{12}}=\text{median\ensuremath{\left\{  \boldsymbol{\hat{\beta}_{3,q}^{12}},\boldsymbol{\hat{\beta}_{4,q}^{12}},\hat{\beta}_{5,q}^{12}\right\} } }$,
$\boldsymbol{\hat{\beta}_{mm,q}^{1}}=\text{median}\left\{ \boldsymbol{\hat{\beta}_{m,q}^{12}},\boldsymbol{\hat{\beta}_{m,q}^{13}},\boldsymbol{\hat{\beta}_{m,q}^{14}},\hat{\beta}_{m,q}^{15}\right\} $.
$\boldsymbol{\hat{\beta}_{mm_{2},q}}=\text{median}\left\{ \boldsymbol{\hat{\beta}_{mm,q}^{1}},\boldsymbol{\hat{\beta}_{mm,q}^{2}},\boldsymbol{\hat{\beta}_{mm,q}^{3}},\boldsymbol{\hat{\beta}_{mm,q}^{4}},\hat{\beta}_{mm,q}^{5}\right\} $. }{\small\par}
\end{table}

Next, for $k_{x}>1$ and one level up in the tree, consider the $l_{2}=\prod_{j=0}^{k_{x}-3}\left(k_{z}-j\right)$
sets of $k_{x}-2$ instruments $\mathcal{L}_{2}=\left\{ L_{2,r_{2}}\right\} _{r_{2}=1}^{l_{2}}$.
Consider the $l_{1}$ sets $\left\{ L_{2,r_{2}},j\right\} _{r_{2}=1,j\in S_{L_{2,r_{2}}}}^{l_{2}}$.
Each set $\left\{ L_{2,r_{2}},j\right\} $ corresponds to a set $L_{1,r_{1}}$.
Therefore, if $L_{2,r_{2}}$ is a set containing valid instruments
only, it follows that $\hat{\boldsymbol{\beta}}_{m}^{\left\{ L_{2,r_{2}},j\right\} }$
is consistent under the conditions stated above iff $j$ is a valid
instrument. Given a set $L_{2,r_{2}}$, we have a total of $k_{z}-k_{x}+2$
estimators $\hat{\boldsymbol{\beta}}_{m}^{\left\{ L_{2,r_{2}},j\right\} }$.
For $k_{\mathcal{V}}>\frac{k_{z}+k_{x}-1}{2}$ it follows that $k_{\mathcal{V}_{S_{L_{2,r_{2}}}}}>\frac{k_{z}-k_{x}+3}{2}>\frac{k_{z}-k_{x}+2}{2}$
for a set $L_{2,r_{2}}$ that contains valid instruments only, where
$k_{\mathcal{V}_{S_{L_{2,r_{2}}}}}$ denotes the number of valid instruments
in $S_{L_{2,r_{2}}}$. Then if $k_{\mathcal{V}}>\frac{k_{z}+k_{x}-1}{2}$
and $L_{2,r_{2}}$is a set of valid instruments it follows that the
median-of-medians estimator 
\[
\hat{\boldsymbol{\beta}}_{mm}^{L_{2,r_{2}}}=\text{median}\left\{ \hat{\boldsymbol{\beta}}_{m}^{\left\{ L_{2,r_{2}},j\right\} }\right\} _{j\in S_{L_{2,r_{2}}}}
\]
is a consistent estimator of $\boldsymbol{\beta}$, and from Propositions
\ref{prop:bmj} and \ref{prop:mm} it follows that $\hat{\boldsymbol{\beta}}_{mm}^{L_{2,r_{2}}}\stackrel{p}{\rightarrow}\boldsymbol{\beta}$
and $\sqrt{\ensuremath{n}}\left(\hat{\boldsymbol{\beta}}_{m}^{L_{2,r_{2}}}-\boldsymbol{\beta}\right)=O_{p}\left(1\right)$.
If $k_{x}=2$ then $\mathcal{L}_{2}=\emptyset$ and the median-of-medians
estimator (\ref{eq:bmm}) is obtained.

For $k_{x}>2$, we repeat the exercise described above and consider
the $l_{3}=\prod_{j=0}^{k_{x}-4}\left(k_{z}-j\right)$ sets of $k_{x}-3$
instruments $\mathcal{L}_{3}=\left\{ L_{3,r_{3}}\right\} _{r_{3}=1}^{l_{3}}$
and the $l_{2}$ sets $\left\{ L_{3,r_{3}},j\right\} _{r_{3}=1,j\in S_{L_{3,r_{3}}}}^{l_{3}}$.
Given a set $L_{3,r_{3}}$, we have a total of $k_{z}-k_{x}+3$ estimators
$\hat{\boldsymbol{\beta}}_{mm}^{\left\{ L_{3,r_{3}},j\right\} }$.
For $k_{\mathcal{V}}>\frac{k_{z}+k_{x}-1}{2}$ it follows that $k_{\mathcal{V}_{S_{L_{3,r_{3}}}}}>\frac{k_{z}-k_{x}+5}{2}>\frac{k_{z}-k_{x}+3}{2}$
for a set $L_{3,r_{3}}$ that contains valid instruments only. This
then implies that more than half of the $\left\{ \hat{\boldsymbol{\beta}}_{mm}^{\left\{ L_{2,r_{3}},j\right\} }\right\} _{j\in S_{L_{3,r_{3}}}}$
are consistent, and hence the median-of-medians-of-medians estimator
\[
\hat{\boldsymbol{\beta}}_{mm_{2}}^{L_{3,r_{3}}}=\text{median}\left\{ \hat{\boldsymbol{\beta}}_{mm}^{\left\{ L_{3,r_{3}},j\right\} }\right\} _{j\in S_{L_{3,r_{3}}}}
\]
is again a consistent estimator of $\boldsymbol{\beta}$ if $k_{\mathcal{V}}>\frac{k_{z}+k_{x}-1}{2}$
and $L_{3,r_{3}}$ is a set of valid instruments. From the proofs
of Propositions \ref{prop:bmj} and \ref{prop:mm} it follows that
$\hat{\boldsymbol{\beta}}_{mm_{2}}^{L_{3,r_{3}}}\stackrel{p}{\rightarrow}\boldsymbol{\beta}$
and $\sqrt{\ensuremath{n}}\left(\hat{\boldsymbol{\beta}}_{mm_{2}}^{L_{3,r_{3}}}-\boldsymbol{\beta}\right)=O_{p}\left(1\right)$.
If $k_{x}=3$, then $\mathcal{L}_{3}=\emptyset$ , resulting in the
median-of-medians-of-medians estimator
\[
\hat{\boldsymbol{\beta}}_{mm_{2}}=\text{median}\left\{ \hat{\boldsymbol{\beta}}_{mm}^{j}\right\} _{j=1}^{k_{z}},
\]
where the subscript 2 indicates that we have taken medians of medians
twice. Table \ref{tab:Illusm2} gives an illustration of this estimator
for $k_{x}=3$ and $k_{z}=5$.

For $k_{x}>3$, we further cascade up the tree considering the $l_{a}=$$\prod_{j=0}^{k_{x}-a-1}\left(k_{z}-j\right)$
sets of $k_{x}-a$ instruments $\mathcal{L}_{a}=\left\{ L_{a,r_{a}}\right\} _{r_{a}=1}^{l_{a}}$,
for $a=4,...,k_{x}$ sequentially. Given consistency of $\hat{\boldsymbol{\beta}}_{mm_{a-2}}^{L_{a-1,r_{a-1}}}$
and given a set $L_{a,r_{a}}$, we have a total of $k_{z}-k_{x}+a$
estimators $\hat{\boldsymbol{\beta}}_{mm_{a-2}}^{\left\{ L_{a,r_{a}},j\right\} }$,
with $j\in S_{L_{a,r_{a}}}$. For $k_{\mathcal{V}}>\frac{k_{z}+k_{x}-1}{2}$
it follows that $k_{\mathcal{V}_{S_{L_{a,r_{a}}}}}>\frac{k_{z}-k_{x}+2a-1}{2}>\frac{k_{z}-k_{x}+a}{2}$
for a set $L_{a,r_{a}}$ that contains valid instruments only. This
then implies that more than half of the $\left\{ \hat{\boldsymbol{\beta}}_{mm_{a-2}}^{\left\{ L_{a,r_{a}},j\right\} }\right\} _{j\in S_{L_{a,r_{a}}}}$
are consistent with the result that
\[
\hat{\boldsymbol{\beta}}_{mm_{a-1}}^{L_{a,r_{a}}}=\text{median}\left\{ \hat{\boldsymbol{\beta}}_{mm_{a-2}}^{\left\{ L_{a,r_{a}},j\right\} }\right\} _{j\in S_{L_{a,r_{a}}}}
\]
are consistent estimators of $\boldsymbol{\beta}$ if $k_{\mathcal{V}}>\frac{k_{z}+k_{x}-1}{2}$
and the set $L_{a,r_{a}}$ contains valid instruments only, for $a=4,\ldots,k_{x}$.
As we have shown consistency of $\hat{\boldsymbol{\beta}}_{mm_{2}}^{L_{3,r_{3}}}=\hat{\boldsymbol{\beta}}_{mm_{a-2}}^{L_{a-1,r_{a-1}}}$
for $a=4$, the results follows for $a=4,\ldots,k_{x}.$

As $\mathcal{L}_{k_{x}}=\emptyset$ we then get the general result
that
\begin{equation}
\hat{\boldsymbol{\beta}}_{mm_{k_{x}-1}}=\text{median}\left\{ \hat{\boldsymbol{\beta}}_{mm_{k_{x}-2}}^{j}\right\} _{j=1}^{k_{z}}\label{eq:bmm2p}
\end{equation}
is a consistent estimator of $\boldsymbol{\beta}$ when $k_{\mathcal{V}}>\frac{k_{z}+k_{x}-1}{2}$.
From the proofs of Propositions \ref{prop:bmj} and \ref{prop:mm}
it follows that $\hat{\boldsymbol{\beta}}_{mm_{k_{x}-1}}\stackrel{p}{\rightarrow}\boldsymbol{\beta}$
and $\sqrt{\ensuremath{n}}\left(\hat{\boldsymbol{\beta}}_{mm_{k_{x}-1}}-\boldsymbol{\beta}\right)=O_{p}\left(1\right)$.
Note that this generalizes the results to all values $k_{x}=1,2,3,....$,
with $\hat{\boldsymbol{\beta}}_{mm_{1}}=\hat{\boldsymbol{\beta}}_{mm}$
and defining $\hat{\boldsymbol{\beta}}_{mm_{0}}:=\hat{\boldsymbol{\beta}}_{m}$
and $\hat{\boldsymbol{\beta}}_{mm_{-1}}^{j}:=\hat{\beta}_{j}$, with
$\hat{\beta_{j}}$ as defined at the beginning of Section \ref{sec:MM}.

We can now summarize the results obtained for general $k_{x}$ in
the following proposition.
\begin{prop}
\label{prop:all} Under Assumptions \ref{assexp}, \ref{clt}, \ref{ass:justrank}
and the generalized majority rule
\begin{equation}
k_{\mathcal{V}}>\frac{k_{z}+k_{x}-1}{2},\label{eq:gmr}
\end{equation}
consider the generalized median-of-medians estimator $\hat{\boldsymbol{\beta}}_{mm_{k_{x}-1}}$
as defined in (\ref{eq:bmm2p}), for $k_{x}=1,2,3,\ldots$. Then $\hat{\boldsymbol{\beta}}_{mm_{k_{x}-1}}\stackrel{p}{\rightarrow}\boldsymbol{\beta}$
and $\sqrt{n}\left(\hat{\boldsymbol{\beta}}_{mm_{k_{x}-1}}-\boldsymbol{\beta}\right)=O_{p}\left(1\right)$. 
\end{prop}

\section{Consistent Selection and Oracle Estimator}

Given the consistent estimator $\hat{\bm{\beta}}_{mm_{k_{x}-1}}$
as defined in (\ref{eq:bmm2p}) we obtain a consistent estimator for
$\bm{\alpha}$ as 
\begin{equation}
\hat{\bm{\alpha}}_{mm_{k_{x}-1}}=\left(\mathbf{Z}^{T}\mathbf{Z}\right)^{-1}\mathbf{Z}^{T}\left(\mathbf{y}-\mathbf{X}\hat{\boldsymbol{\beta}}_{mm_{k_{x}-1}}\right).\label{alpha-mm}
\end{equation}
From the properties of $\hat{\bm{\beta}}_{mm_{k_{x}-1}}$ as given
in Proposition \ref{prop:all}, $\hat{\bm{\alpha}}_{mm_{k_{x}-1}}$
satisfies the conditions of Assumption \ref{ass:alphahat}, $\hat{\bm{\alpha}}_{mm_{k_{x}-1}}\stackrel{p}{\rightarrow}\boldsymbol{\alpha}$
and $\sqrt{n}\left(\hat{\bm{\alpha}}_{mm_{k_{x}-1}}-\boldsymbol{\alpha}\right)=O_{p}\left(1\right)$.
Therefore, from Theorem 2 and Remark 1 in \citet{Zou2006adaptive},
it follows that the adaptive Lasso estimator $\hat{\bm{\alpha}}_{ad}$
that uses $\hat{\bm{\alpha}}_{mm_{k_{x}-1}}$ as the initial consistent
estimator satisfies consistency of selection and oracle properties
as stated in the following proposition.
\begin{prop}
\label{prop:ador} Under the conditions of Proposition \ref{prop:all}
and Assumption \ref{ass:lambdan} for $\lambda_{n}$, let $\hat{\bm{\alpha}}_{mm_{k_{x}-1}}$
as defined in (\ref{alpha-mm}) be the initial consistent estimator
in the adaptive Lasso criterion (\ref{alasso selection}). Let $\hat{\mathcal{\mathcal{A}}}_{ad}=\left\{ j:\hat{\alpha}_{ad,j}\neq0\right\} $.
Then the adaptive Lasso estimator $\hat{\bm{\alpha}}_{ad}$ satisfies
$\lim_{n\rightarrow\infty}P\left(\hat{\mathcal{\mathcal{A}}}_{ad}=\mathcal{A}\right)=1$
and the limiting normal distribution of $\sqrt{n}\left(\hat{\bm{\alpha}}_{ad,\mathcal{A}}-\boldsymbol{\alpha}_{\mathcal{A}}\right)$
is that of the oracle 2sls estimator $\hat{\bm{\alpha}}_{2sls}^{or}$
as defined in (\ref{eq:alphaor}) with the limiting distribution as
given in (\ref{eq:thetaorlim}).
\end{prop}
Similar to \citet{KangetalJASA2016} and \citet{WindmeijeretalJASA2019},
the adaptive Lasso estimator for $\bm{\beta}$ is obtained as 
\begin{equation}
\hat{\bm{\beta}}_{ad}=\left(\hat{\mathbf{X}}^{T}\hat{\mathbf{X}}\right){}^{-1}\hat{\mathbf{X}}^{T}\left(\mathbf{y}-\mathbf{Z}\hat{\bm{\alpha}}_{ad}\right).\label{alassoestimator}
\end{equation}
From the results of Proposition \ref{prop:ador} it follows that the
limiting distribution of $\hat{\bm{\beta}}_{ad}$ is that of the oracle
2sls estimator, as stated in the next corollary.
\begin{cor}
\label{cor:betaad} Let $\hat{\bm{\beta}}_{ad}$ as defined in (\ref{alassoestimator}).
Under the conditions of Proposition \ref{prop:ador} the limiting
normal distribution of $\sqrt{n}\left(\hat{\bm{\beta}}_{ad}-\boldsymbol{\beta}\right)$
is that of the oracle 2sls estimator $\hat{\bm{\beta}}_{2sls}^{or}$
as defined in (\ref{eq:betaor}), with the limiting distribution as
given in (\ref{eq:thetaorlim}).
\end{cor}
As an alternative to obtaining the causal estimator directly from
the adaptive Lasso as in (\ref{alassoestimator}), we can also estimate
$\bm{\beta}$ by post-selection 2sls using the estimated set of invalid
instruments $\hat{\mathcal{A}}_{ad}$ in the following specification:
\begin{equation}
\mathbf{y}=\mathbf{X}\bm{\beta}+\mathbf{Z}_{\hat{\mathcal{A}}_{ad}}\bm{\alpha}_{\hat{\mathcal{A}}_{ad}}+\mathbf{u},\label{post-2sls}
\end{equation}
using $\mathbf{Z}_{\hat{\mathcal{V}}_{ad}}$ as the set of valid instruments,
where $\hat{\mathcal{V}}_{ad}=\left\{ j:\hat{\alpha}_{ad,j}=0\right\} $
. The next proposition states the oracle properties of the post-selection
2sls estimator in model specification (\ref{post-2sls}). The proof
follows directly from Theorem 2 in Guo et al. (2018) as, under the
stated conditions, $\lim_{n\rightarrow\infty}P(\hat{\mathcal{V}}_{ad}=\mathcal{V})=1$.
\begin{prop}
\label{prop:2slsp} Let $\hat{\bm{\beta}}_{2sls,p}$ be the post-selection
2sls estimator of $\boldsymbol{\beta}$ in model (\ref{post-2sls}),
which is given by 
\[
\hat{\bm{\beta}}_{2sls,p}=\left(\hat{\mathbf{X}}^{\prime}\mathbf{M}_{\mathbf{Z}_{\hat{\mathcal{A}}_{ad}}}\hat{\mathbf{X}}\right)^{-1}\hat{\mathbf{X}}^{\prime}\mathbf{M}_{\mathbf{Z}_{\hat{\mathcal{A}}_{ad}}}\mathbf{y}.
\]
Under the conditions of Corollary \ref{cor:betaad}, it follows that
the limiting normal distribution of $\sqrt{n}\left(\hat{\bm{\beta}}_{2sls,p}-\bm{\beta}\right)$
is that of the of the oracle 2sls estimator $\hat{\bm{\beta}}_{2sls}^{or}$
as defined in (\ref{eq:betaor}), with the limiting distribution as
given in (\ref{eq:thetaorlim}).
\end{prop}

\subsection{Downward Testing Procedure}

\label{sec:downward} Consistent IV selection using the adaptive Lasso
depends on the choice of the tuning parameter $\lambda_{n}$ which
controls the strength of penalization. While $\lambda_{n}$ needs
to satisfy the theoretical conditions of Assumption \ref{ass:lambdan},
$n^{\frac{1-\nu}{2}}\lambda_{n}\rightarrow\infty$, $\lambda_{n}=o(\sqrt{n})$,
it can be challenging to pick a specific value of $\lambda_{n}$ for
a given sample. A common practice of choosing the tuning parameter
is k-fold cross-validation. However, it is well known that cross-validation
works better for prediction rather than model selection \citep{buhlmann2011statistics},
and cross-validation almost always results in inconsistent variable
selection, as stated in \citet{chand2012tuning}.

As an alternative, and similar to \citet{WindmeijeretalJASA2019}
and \citet{windmeijer2021confidence}, we combine the adaptive Lasso
with the downward testing procedure for moment selection as proposed
by \citet{Andrews1999}, which uses the Sargan test statistic as the
selection criterion, as defined in (\ref{eq:SarganS}). A crude downward
testing procedure starts with the model $\mathbf{y}=\mathbf{X}\boldsymbol{\beta}+\mathbf{u}$,
treating all $k_{z}$ instruments as valid. If the Sargan test rejects
the model, then the procedure moves to models with $k_{z}-1$ treated
as valid instruments and tests all such models $\mathbf{y}=\mathbf{X}\boldsymbol{\beta}+\mathbf{z}_{j}\alpha_{j}+\mathbf{u}_{j}$,
$j=1,\ldots,k_{z}$. If the Sargan test rejects them all, then it
moves to evaluate all $\binom{k_{z}}{2}$ models with $k_{z}-2$ instruments
treated as valid, and so on, until it finds a model that is not rejected
by the Sargan test. This procedure can become computationally infeasible
since for each number of instruments, $k_{z},k_{z}-1,...$, we need
to exhaustively test models corresponding to all possible combinations
of instruments.

The adaptive Lasso can mitigate the computational challenges in the
downward testing procedure. When the adaptive Lasso is implemented
using the Least-Angle Regression (LARS) algorithm \citep{Efron2004Least},
it generates a selection path starting with a model with $k_{z}$
valid instruments, and, for each LARS step, the number of instruments
treated as valid decreases by one. This means that, for each number
of instruments treated as valid, $k_{z},k_{z}-1,...$, we only need
to evaluate a single model, i.e.\ the one on the LARS selection path.
Given the consistency of selection and oracle results of the adaptive
Lasso estimator as given in Proposition \ref{prop:ador}, the oracle
model lies on this path in large samples. Given the properties of
the Sargan test as described in Section \ref{subsec:IV-Estimation}
and the adjusted majority rule requirement as given in Assumption
\ref{ass:jmajority}, it follows that selecting the first model on
this LARS path that does not reject the Sargan test is a consistent
selection rule, when for a model with $k_{inv}$ instruments selected
as invalid, the critical value $\zeta_{n,k_{z}-k_{x}-k_{inv}}$ used
for the $\chi_{k_{z}-k_{x}-k_{inv}}^{2}$ distribution satisfies 
\begin{equation}
\zeta_{n,k_{z}-k_{x}-k_{inv}}\rightarrow\infty\text{ for }n\rightarrow\infty\text{, and \,}\zeta_{n,k_{z}-k_{x}-k_{inv}}=o\left(n\right),\label{eq:ch2critical}
\end{equation}
see \citet{Andrews1999}. In practice, following \citet{WindmeijeretalJASA2019}
and \citet{windmeijer2021confidence}, instead of a critical value
$\zeta_{n,k_{z}-k_{x}-k_{inv}}$ for the Sargan test, we use a p-value
$p_{n}$. If $p_{n}$ satisfies $\lim_{n\rightarrow\infty}p_{n}=0$
and $\log\left(p_{n}\right)=o\left(n\right)$, then condition (\ref{eq:ch2critical})
is satisfied. As in \citet{WindmeijeretalJASA2019} and \citet{windmeijer2021confidence},
for a given sample, we set $p_{n}=0.1/\log\left(n\right)$, as suggested
by \citet{BellonietalEcta2012}. This procedure leads to consistent
selection and oracle properties of the post-selection 2sls estimator
as detailed in Proposition \ref{prop:2slsp}. 

\section{Instrument Relevance}

\label{sec: extension}

In the previous sections, we maintained Assumption \ref{ass:justrank},
requiring that all just identified models identify all parameters
in $\boldsymbol{\beta}$. However, in practical applications, it may
well be the case that a given instrument is not relevant for all endogenous
exposure variables. In this case, some just-identifying combination
of instruments would violate the full rank assumption. In practice,
one could test for underidentification, as described in e.g.\ \citet{WindmeijerJoE2021},
and discard just identified estimates where the test for underidentification
fails to reject, similar to the first-stage hard-thresholding method
of \citet{GuoetalJRSSB2018}. However, it may be difficult then to
establish an adjusted majority rule as in Assumption \ref{ass:jmajority}
and is the subject of future research.

Instead, we consider here the case, as in our application, where the
the relevance of the instruments with respect to each endogenous exposure
variable is known. In our application of Mendelian randomization,
the potential instruments are genetic markers, which are identified
from GWAS, and hence it is known from these studies which genetic
marker is relevant for which exposure variable. In Mendelian randomization
studies, the genetic markers are also independently distributed. We
show here how to obtain the consistent median-of-medians estimator
that incorporates this information. For ease of exposition and in
line with our application, we focus here on the $k_{x}=2$ case.

We first consider the case where each instrument is relevant only
for one of the exposure variables. For the $k_{x}=2$ case, the first-stage
linear specification can then be written as
\[
\mathbf{X}=\left[\mathbf{x}_{1}\,\,\mathbf{x}_{2}\right]=\left[\mathbf{Z}_{1}\,\,\mathbf{Z}_{2}\right]\left[\begin{array}{cc}
\boldsymbol{\pi}_{1} & \mathbf{0}\\
\mathbf{0} & \boldsymbol{\pi}_{2}
\end{array}\right]+\mathbf{E},
\]
where $\mathbf{Z}_{1}$ is the $n\times k_{1}$ matrix of instruments
relevant for $\mathbf{x}_{1}$ and $\mathbf{Z}_{2}$ is the $n\times k_{2}$
matrix of instruments relevant for $\mathbf{x}_{2}$. When instruments
are independent, as generally the case in Mendelian randomization
studies, any just-identifying pair of them can identify the parameter
vector $\boldsymbol{\beta}=\left(\beta_{1}\,\,\beta_{2}\right)^{T}$
only if it combines one instrument from $\mathbf{Z}_{1}$ with one
instrument from $\mathbf{Z}_{2}$. Hence there are now $k_{1}\times k_{2}$
sets of just-identifying instruments that are relevant for both exposure
variables. Let $k_{\mathcal{V}_{1}}$ and $k_{\mathcal{V}_{2}}$ denote
the number of valid instruments in $\mathbf{Z}_{1}$ and $\mathbf{Z}_{2}$
respectively. Then there are $k_{\mathcal{V}_{1}}\times k_{\mathcal{V}_{2}}$
pairs of instruments where the instruments are both valid and hence
for the naive median estimator to be consistent, the condition that
$k_{\mathcal{V}_{1}}\times k_{\mathcal{V}_{2}}>k_{1}\times k_{2}/2$
needs to hold.

For the median-of medians estimator to be consistent, we now require
that $k_{\mathcal{V}_{1}}>\frac{k_{1}}{2}$ and $k_{\mathcal{V}_{2}}>\frac{k_{2}}{2}$,
or the standard majority rule holds for each set. This can be shown
as follows. Let the indices of the instruments relevant for $\mathbf{x}_{1}$
be $S_{1}=\left\{ 1,2,\ldots,k_{1}\right\} $, and those for $\mathbf{x}_{2}$
be $S_{2}=\left\{ k_{1}+1,k_{1}+2,\ldots,k_{z}\right\} $, where $k_{z}=k_{1}+k_{2}$.
We then have the just identified IV estimators $\hat{\boldsymbol{\beta}}_{s}^{j}$,
with, when $j\in S_{1}$, $s\in S_{2}$ and vice versa. For each element
$\beta_{q}$ in $\boldsymbol{\beta}$, $q=1,2$, we have for each
instrument $j$ a vector of estimators $\hat{\boldsymbol{\beta}}_{q}^{j}=\left(\hat{\beta}_{s,q}^{j}\right)$.
This is a $k_{2}$-vector if $j\in S_{1}$ and a $k_{1}$-vector is
$j\in S_{2}$. Let $\hat{\beta}_{m,q}^{j}=\text{median}\left(\hat{\boldsymbol{\beta}}_{q}^{j}\right)$.
Then $\hat{\beta}_{m,q}^{j}$ is consistent if $j$ is a valid instrument
and if $j\in S_{1}$, $k_{\mathcal{V}_{2}}>\frac{k_{2}}{2}$, or if
$j\in S_{2}$, $k_{\mathcal{V}_{1}}>\frac{k_{1}}{2}$. There are then
$k_{\mathcal{V}_{1}}+k_{\mathcal{V}_{2}}>\frac{k_{z}}{2}$ consistent
estimators in $\left\{ \hat{\beta}_{m,q}^{j}\right\} _{j=1}^{k_{z}}$
and hence $\hat{\beta}_{mm,q}=\text{median}\left\{ \hat{\beta}_{m,q}^{j}\right\} _{j=1}^{k_{z}}$
is a consistent estimator of $\beta_{q}$, for $q=1,2$.

This result is illustrated in Table \ref{tab: mmillusblock} with
an example where $S_{1}=\left\{ 1,2,3,4\right\} $, and $S_{2}=\left\{ 5,6,7\right\} $.
Valid instruments are $\mathcal{V}_{1}=\left\{ 2,3,4\right\} $, and
$\mathcal{V}_{2}=\left\{ 6,7\right\} $, and so $\mathcal{A}_{1}=\left\{ 1\right\} $
and $\mathcal{A}_{2}=\left\{ 5\right\} $. Therefore the individual
majority rule for each set holds, and no more instruments can be invalid.
Although the total number of two instruments allowed to be invalid
is the same here as in the example of Table \ref{tab: mmillus}, it
is clear that they cannot be both in the set that is relevant for
one of the exposure variables.

\begin{table}[H]
\begin{centering}
\caption{\label{tab: mmillusblock} Illustration of median-of-medians estimator
with block structure relevance of instruments.}
\par\end{centering}
\medskip{}

\begin{centering}
\begin{tabular}{ccccccccc}
\toprule 
Instruments & $1$ & $\boldsymbol{2}$ & $\boldsymbol{3}$ & $\boldsymbol{4}$ & $5$ & $\boldsymbol{6}$ & $\boldsymbol{7}$ & \tabularnewline
\midrule 
$1$ &  &  &  &  & $\hat{\beta}_{1,q}^{5}$ & $\hat{\beta}_{1,q}^{6}$ & $\hat{\beta}_{1,q}^{7}$ & \tabularnewline
$\boldsymbol{2}$ &  &  &  &  & $\hat{\beta}_{2,q}^{5}$ & $\boldsymbol{\hat{\beta}_{2,q}^{6}}$ & $\boldsymbol{\hat{\beta}_{2,q}^{7}}$ & \tabularnewline
$\boldsymbol{3}$ &  &  &  &  & $\hat{\beta}_{3,q}^{5}$ & $\boldsymbol{\hat{\beta}_{3,q}^{6}}$ & $\boldsymbol{\hat{\beta}_{3,q}^{7}}$ & \tabularnewline
$\boldsymbol{4}$ &  &  &  &  & $\hat{\beta}_{4,q}^{5}$ & $\boldsymbol{\hat{\beta}_{4,q}^{6}}$ & $\boldsymbol{\hat{\beta}_{4,q}^{7}}$ & \tabularnewline
$5$ & $\hat{\beta}_{5,q}^{1}$ & $\hat{\beta}_{5,q}^{2}$ & $\hat{\beta}_{5,q}^{3}$ & $\hat{\beta}_{5,q}^{2}$ &  &  &  & \tabularnewline
$\boldsymbol{6}$ & $\hat{\beta}_{6,q}^{1}$ & $\boldsymbol{\hat{\beta}_{6,q}^{2}}$ & $\boldsymbol{\hat{\beta}_{6,q}^{3}}$ & $\boldsymbol{\hat{\beta}_{6,q}^{4}}$ &  &  &  & \tabularnewline
$\boldsymbol{7}$ & $\hat{\beta}_{7,q}^{1}$ & $\boldsymbol{\hat{\beta}_{7,q}^{2}}$ & $\boldsymbol{\hat{\beta}_{7,q}^{3}}$ & $\boldsymbol{\hat{\beta}_{7,q}^{4}}$ &  &  &  & \tabularnewline
\midrule 
median & $\hat{\beta}_{m,q}^{1}$ & $\boldsymbol{\hat{\beta}_{m,q}^{2}}$ & $\boldsymbol{\hat{\beta}_{m,q}^{3}}$ & $\boldsymbol{\hat{\beta}_{m,q}^{4}}$ & $\hat{\beta}_{m,q}^{5}$ & $\boldsymbol{\hat{\beta}_{m,q}^{6}}$ & $\boldsymbol{\hat{\beta}_{m,q}^{7}}$ & $\boldsymbol{\hat{\beta}_{mm,q}}$\tabularnewline
\bottomrule
\end{tabular}
\par\end{centering}
\centering{}{\small{}Notes: $k_{x}=2$, $S_{1}=\left\{ 1,2,3,4\right\} $,
$S_{2}=\left\{ 5,6,7\right\} $,$\mathcal{V}=\left\{ 2,3,4,6,7\right\} $,
$\mathcal{A}=\left\{ 1,5\right\} $. Valid instruments and consistent
estimators are displayed in boldface.}{\small\par}
\end{table}

These conditions can change if there is some overlap between the two
groups, for example if $S_{1}=\left\{ 1,2,3,4,5\right\} $ and $S_{2}=\left\{ 5,6,7\right\} $,
both instruments $1$ and $2$ can be invalid, as illustrated in Table
\ref{tab: mmillusover}, as the majority in $S_{1}$ is valid. This
is not the case if instead $S_{1}=\left\{ 1,2,3,4\right\} $ and $S_{2}=\left\{ 4,5,6,7\right\} $.

\begin{table}[H]
\begin{centering}
\caption{\label{tab: mmillusover} Illustration of median-of-medians estimator
with block structure relevance of instruments with overlap.}
\par\end{centering}
\medskip{}

\begin{centering}
\begin{tabular}{ccccccccc}
\toprule 
Instruments & $1$ & $2$ & $\boldsymbol{3}$ & $\boldsymbol{4}$ & $\boldsymbol{5}$ & $\boldsymbol{6}$ & $\boldsymbol{7}$ & \tabularnewline
\midrule 
$1$ &  &  &  &  & $\hat{\beta}_{1,q}^{5}$ & $\hat{\beta}_{1,q}^{6}$ & $\hat{\beta}_{1,q}^{7}$ & \tabularnewline
$2$ &  &  &  &  & $\hat{\beta}_{2,q}^{5}$ & $\hat{\beta}_{2,q}^{6}$ & $\hat{\beta}_{2,q}^{7}$ & \tabularnewline
$\boldsymbol{3}$ &  &  &  &  & $\boldsymbol{\hat{\beta}_{3,q}^{5}}$ & $\boldsymbol{\hat{\beta}_{4,q}^{7}}$ & $\boldsymbol{\hat{\beta}_{3,q}^{7}}$ & \tabularnewline
$\boldsymbol{4}$ &  &  &  &  & $\boldsymbol{\hat{\beta}_{4,q}^{5}}$ & $\boldsymbol{\hat{\beta}_{3,q}^{7}}$ & $\boldsymbol{\hat{\beta}_{4,q}^{7}}$ & \tabularnewline
$\boldsymbol{5}$ & $\hat{\beta}_{5,q}^{1}$ & $\hat{\beta}_{5,q}^{2}$ & $\boldsymbol{\hat{\beta}_{5,q}^{3}}$ & $\boldsymbol{\hat{\beta}_{5,q}^{4}}$ &  & $\boldsymbol{\hat{\beta}_{5,q}^{6}}$ & $\boldsymbol{\hat{\beta}_{5,q}^{6}}$ & \tabularnewline
$\boldsymbol{6}$ & $\hat{\beta}_{6,q}^{1}$ & $\hat{\beta}_{5,q}^{2}$ & $\boldsymbol{\hat{\beta}_{4,q}^{3}}$ & $\boldsymbol{\hat{\beta}_{6,q}^{4}}$ & $\boldsymbol{\hat{\beta}_{6,q}^{5}}$ &  &  & \tabularnewline
$\boldsymbol{7}$ & $\hat{\beta}_{7,q}^{1}$ & $\hat{\beta}_{5,q}^{2}$ & $\boldsymbol{\hat{\beta}_{7,q}^{3}}$ & $\boldsymbol{\hat{\beta}_{7,q}^{4}}$ & $\boldsymbol{\hat{\beta}_{7,q}^{5}}$ &  &  & \tabularnewline
\midrule 
median & $\hat{\beta}_{m,q}^{1}$ & $\hat{\beta}_{m,q}^{2}$ & $\boldsymbol{\hat{\beta}_{m,q}^{3}}$ & $\boldsymbol{\hat{\beta}_{m,q}^{4}}$ & $\boldsymbol{\hat{\beta}_{m,q}^{5}}$ & $\boldsymbol{\hat{\beta}_{m,q}^{6}}$ & $\boldsymbol{\hat{\beta}_{m,q}^{7}}$ & $\boldsymbol{\hat{\beta}_{mm,q}}$\tabularnewline
\bottomrule
\end{tabular}
\par\end{centering}
\centering{}{\small{}Notes: $k_{x}=2$, $S_{1}=\left\{ 1,2,3,4,5\right\} $,
$S_{2}=\left\{ 5,6,7\right\} $,$\mathcal{V}=\left\{ 3,4,5,6,7\right\} $,
$\mathcal{A}=\left\{ 1,2\right\} $. Valid instruments and consistent
estimators are displayed in boldface.}{\small\par}
\end{table}

\section{Monte Carlo Simulations}

\label{sec: simulation}

We conduct Monte Carlo simulations to evaluate the performance of
our method in two settings. In the first setting, all instruments
are relevant for both of the endogenous variables, while in the other
setting, each instruemnt is relevent for one exposure only. We run
the simulations for $1,000$ replications, and we implement the adaptive
Lasso using the \texttt{Lars} package \citep{lars} in \texttt{R}.
We set $k_{x}=2$ and generate the data from
\begin{align*}
\mathbf{y} & =\mathbf{X}\boldsymbol{\beta}+\mathbf{Z}\boldsymbol{\alpha}+\mathbf{u}\\
\mathbf{x_{1}} & =\mathbf{Z}\boldsymbol{\pi}_{1}+\mathbf{e}_{1}\\
\mathbf{x_{2}} & =\mathbf{Z}\boldsymbol{\pi}_{2}+\mathbf{e}_{2},
\end{align*}
where 
\[
\left(\begin{array}{c}
U_{i}\\
E_{1i}\\
E_{2i}
\end{array}\right)\sim N\left(\left(\begin{array}{c}
0\\
0\\
0
\end{array}\right),\left(\begin{array}{ccc}
1 & \rho_{1} & \rho_{2}\\
\rho_{1} & 1 & 0\\
\rho_{2} & 0 & 1
\end{array}\right)\right);
\]
\[
\mathbf{Z}_{i}\sim N\left(\mathbf{0},\bm{\Sigma}_{z}\right);
\]
with $\bm{\beta}=(0.3,0.6)^{T}$; $k_{z}=21$; $\rho_{1}=0.25$, $\rho_{2}=0.3$;
$k_{\mathcal{V}}=12$, $k_{\mathcal{A}}=9$, $\bm{\alpha}=0.4\left(\bm{\iota}_{9}^{T},\mathbf{0}_{12}^{T}\right)^{T}$,
similar to the simulation setup in \citet{windmeijer2021confidence}.
We generate the elements of $\bm{\pi}_{1}$ and $\boldsymbol{\pi}_{2}$
from a uniform distribution on the interval $[1.5,2.5]$, and we set
the elements of $\mathbf{\boldsymbol{\Sigma}}_{z}$ to $\Sigma_{z,jk}=0.5^{|j-k|}$,
$j,k=1,\ldots.,k_{z}$. In this setup, all the instruments are relevant
for both endogenous variables, and both the pair-wise full rank Assumption
\ref{ass:justrank} and the majority Assumption \ref{ass:jmajority}
are satisfied.

Table~\ref{tab:mcresults} presents the IV selection and estimation
results of the adaptive Lasso method. The first two columns of Table
\ref{tab:mcresults} report statistics related to estimation, and
in both of these columns, we average the statistics over the two entries
in $\bm{\beta}$. Column~(1) presents the averaged median absolute
error (MAE), while Column~(2) shows the averaged standard deviation
(SD). The remaining three columns in Table~\ref{tab:mcresults} report
statistics related to IV selection. Column~(3) reports the average
number of instruments selected as invalid, Column~(4) the frequency
with which all invalid instruments have been selected as invalid,
and Column~(5) the frequency with which the oracle model has been
selected. The three panels in Table~\ref{tab:mcresults} correspond
to the sample sizes $n=500,1000,2000$. In each panel, the first row,
denoted ``Oracle 2sls'', shows the results for the oracle 2sls estimator,
which is the 2sls estimator that uses the true set of valid instruments,
while it controls for the remaining invalid ones. The second row,
denoted ``Naive 2sls'', reports the results for the 2sls estimator
that considers all candidate instruments to be valid. The third row,
denoted $\hat{\bm{\beta}}_{mm}$, shows the results for the median-in-medians
estimator, as defined in (\ref{eq:bmm}). The fourth to eighth rows
report the results for the adaptive Lasso estimators, including the
``Post-ALasso'' estimators that are the post-selection 2sls estimators
that use the instruments selected as valid, and include the instruments
selected as invalid as control variables. In rows 3-7 we present results
for the adaptive Lasso estimators using two different types of ten-fold
cross-validation. First, denoted with the ``$cv$'' subscript, the
results use the tuning parameter that gives the minimum cross-validation
Sargan statistics. Second, denoted with the ``$cvse$'' subscript
uses the tuning parameter chosen by the one-standard-error rule, see
\citet{KangetalJASA2016} for details.

\begin{table}
\begin{centering}
\caption{\label{tab:mcresults} Simulation Results}
\par\end{centering}
\medskip{}

\begin{centering}
\begin{tabular}{lccccc}
\toprule 
 & MAE & SD & \# invalid & p allinv & p oracle\tabularnewline
 & (1) & (2) & (3) & (4) & (5)\tabularnewline
\midrule 
\multicolumn{2}{l}{{\footnotesize{}$n=500$}} &  &  &  & \tabularnewline
\hspace{1em} Oracle 2sls & 0.0624 & 0.0912 & 9 & 1 & 1\tabularnewline
\hspace{1em} Naive 2sls & 0.2856 & 0.2685 & 0 & 0 & 0\tabularnewline
\hspace{1em} $\hat{\bm{\beta}}_{mm}$ & 0.1263 & 0.1517 &  &  & \tabularnewline
\hspace{1em} ALasso$_{cv}$ & 0.3460 & 0.4212 & 8.095 & 0.440 & 0.440\tabularnewline
\hspace{1em} Post-ALasso$_{cv}$ & 0.1295 & 0.3060 &  &  & \tabularnewline
\hspace{1em} ALasso$_{cvse}$ & 0.4393 & 0.4542 & 6.908 & 0.115 & 0.115\tabularnewline
\hspace{1em} Post-ALasso$_{cvse}$ & 0.2912 & 0.4368 &  &  & \tabularnewline
\hspace{1em} Post-ALasso$_{Sar}$ & 0.0853 & 0.1446 & 9.09 & 0.987 & 0.947\tabularnewline
\midrule 
\addlinespace
{\footnotesize{}$n=1,000$} &  &  &  &  & \tabularnewline
\hspace{1em} Oracle 2sls & 0.0439 & 0.0681 & 9 & 1 & 1\tabularnewline
\hspace{1em} Naive 2sls & 0.2889 & 0.2037 & 0 & 0 & 0\tabularnewline
\hspace{1em} $\hat{\bm{\beta}}_{mm}$ & 0.0892 & 0.1268 &  &  & \tabularnewline
\hspace{1em} ALasso$_{cv}$ & 0.2047 & 0.2843 & 8.857 & 0.882 & 0.882\tabularnewline
\hspace{1em} Post-ALasso$_{cv}$ & 0.0513 & 0.1627 &  &  & \tabularnewline
\hspace{1em} ALasso$_{cvse}$ & 0.2895 & 0.3446 & 8.509 & 0.634 & 0.634\tabularnewline
\hspace{1em} Post-ALasso$_{cvse}$ & 0.0716 & 0.2332 &  &  & \tabularnewline
\hspace{1em} Post-ALasso$_{Sar}$ & 0.0544 & 0.0690 & 9.02 & 1 & 0.983\tabularnewline
\midrule 
\multicolumn{2}{l}{{\footnotesize{}$n=2,000$}} &  &  &  & \tabularnewline
\hspace{1em} Oracle 2sls & 0.0305 & 0.0473 & 9 & 1 & 1\tabularnewline
\hspace{1em} Naive 2sls & 0.2796 & 0.1448 & 0 & 0 & 0\tabularnewline
\hspace{1em} $\hat{\bm{\beta}}_{mm}$ & 0.0618 & 0.0941 &  &  & \tabularnewline
\hspace{1em} ALasso$_{cv}$ & 0.1341 & 0.1795 & 8.991 & 0.992 & 0.992\tabularnewline
\hspace{1em} Post-ALasso$_{cv}$ & 0.0307 & 0.0574 &  &  & \tabularnewline
\hspace{1em} ALasso$_{cvse}$ & 0.1753 & 0.2244 & 8.949 & 0.956 & 0.956\tabularnewline
\hspace{1em} Post-ALasso$_{cvse}$ & 0.0319 & 0.0881 &  &  & \tabularnewline
\hspace{1em} Post-ALasso$_{Sar}$ & 0.0375 & 0.0478 & 9.018 & 1 & 0.984\tabularnewline
\bottomrule
\end{tabular}
\par\end{centering}
\medskip{}

{\footnotesize{}Notes: The reported statistics include median absolute
error, standard deviation, number of IVs selected as invalid, frequency
with which all invalid IVs have been selected as invalid, and frequency
with which oracle model has been selected. The simulations are based
on $1,000$ repetitions.}{\footnotesize\par}
\end{table}

In terms of IV selection, in all three sample sizes, the $cv$-procedure
dominates the $cvse$-procedure, especially for the smallest sample
with $n=500$. Both methods improve as the sample size increases.
The frequencies of selecting the oracle model are both almost equal
to 1 at $n=2,000$ with 0.992 for $cv$ and 0.956 for $cvse$. In
line with the selection performance, the post-selection 2sls estimates
are close to the oracle model at $n=2,000$. For all three sample
sizes, the post-selection 2sls estimators outperform the adaptive
Lasso estimators.

The eighth row, denoted ``Post-ALasso$_{Sar}$'' displays the results
for the post-selection 2sls estimator using the downward testing procedure
described in Section \ref{sec:downward}. We see that this estimation
procedure outperforms the methods based on cross-validation, in particular
at the smaller sample sizes. For example, we find that it reaches
a high frequency, $0.947$, of selecting the oracle model even at
the smallest sample size $n=500$, a much higher frequency than those
found for the $cv$ ($0.440$) and $cvse$ ($0.115$) methods.

Next, we consider the case where the sets of instruments for $\mathbf{x}_{1}$
and $\mathbf{x}_{2}$ are separate, such that no instrument is relevant
for both endogenous variables. We set $\bm{\pi}_{1}=(\bm{\gamma}_{1}^{T},\mathbf{0}_{11}^{T})^{T}$
and $\bm{\pi}_{2}=(\mathbf{0}_{10}^{T},\bm{\gamma}_{2}^{T})^{T}$,
where $\bm{\gamma}_{1}$ has length $k_{1}=10$ and $\bm{\gamma}_{_{2}}$
has length $k_{2}=11$. We let $\bm{\alpha}=(\bm{\iota}_{4}^{T},\mathbf{0}_{6}^{T},\bm{\iota}_{5}^{T},\mathbf{0}_{6}^{T})^{T}$
such that $k_{\mathcal{V}_{1}}=6$, $k_{\mathcal{A}_{1}}=4$ and $k_{\mathcal{V}_{2}}=6$,
$k_{\mathcal{A}_{1}}=5$. All the other parameters are identical to
the previous simulation design. Again, the necessary and sufficient
majority assumption for consistency of the median-of-medians estimator,
$k_{\mathcal{V}_{1}}>\frac{k_{z_{1}}}{2}$ and $k_{\mathcal{V}_{2}}>\frac{k_{z_{2}}}{2}$,
is satisfied.

Table \ref{tab:mcresblock} reports the results for the oracle and
naive 2sls estimators, the median-of-medians estimator and the Post-ALasso$_{Sar}$
estimator based on the downward testing procedure, both with and without
the block structure to obtain the median-of-medians estimator as discussed
in Section \ref{sec: extension}. The table reports the same statistics
as earlier, and the panels correspond again to the sample sizes $n=500,1000,2000$.
For all three sample sizes, results imposing the block structure dominate
the ones without. It is evident that $\hat{\bm{\beta}}_{mm}$ has
a larger MAE and SD than $\hat{\bm{\beta}}_{mm,block}$ and the frequencies
of selecting the oracle model when imposing the block structure are
larger than those without imposing the block structure. Here the selection
performance without imposing the block structure does actually not
improve with the sample size, and the frequencies of selecting the
oracle model remains at around $0.5$. For the method imposing the
block structure, the oracle selection frequencies are very close to
1, even at $n=500$ ($0.971$).

\begin{table}[H]
\begin{centering}
\caption{\label{tab:mcresblock} Simulation results, separate sets of instruments
for each exposure variable}
\par\end{centering}
\medskip{}

\begin{centering}
\begin{tabular}{lccccc}
\toprule 
 & MAE & SD & \# invalid & p allinv & p oracle\tabularnewline
 & (1) & (2) & (3) & (4) & (5)\tabularnewline
\midrule 
\multicolumn{2}{l}{{\footnotesize{}$n=500$}} &  &  &  & \tabularnewline
\hspace{1em} Oracle 2sls & 0.0109 & 0.0161 & 9 & 1 & 1\tabularnewline
\hspace{1em} Naive 2sls & 0.3285 & 0.0209 & 0 & 0 & 0\tabularnewline
\hspace{1em} $\hat{\bm{\beta}}_{mm}$ & 0.1124 & 0.1472 &  &  & \tabularnewline
\hspace{1em} Post-ALasso$_{Sar}$ & 0.0159 & 0.1779 & 10.241 & 0.791 & 0.515\tabularnewline
\hspace{1em} $\hat{\bm{\beta}}_{mm,block}$ & 0.0839 & 0.0394 &  &  & \tabularnewline
\hspace{1em} Post-ALasso$_{Sar,block}$ & 0.0111 & 0.0192 & 9.044 & 0.999 & 0.971\tabularnewline
\midrule 
\multicolumn{2}{l}{{\footnotesize{}$n=1,000$}} &  &  &  & \tabularnewline
\hspace{1em} Oracle 2sls & 0.0075 & 0.0111 & 9 & 1 & 1\tabularnewline
\hspace{1em} Naive 2sls & 0.3288 & 0.0149 & 0 & 0 & 0\tabularnewline
\hspace{1em} $\hat{\bm{\beta}}_{mm}$ & 0.0892 & 0.1629 &  &  & \tabularnewline
\hspace{1em} Post-ALasso$_{Sar}$ & 0.0102 & 0.2413 & 10.052 & 0.786 & 0.565\tabularnewline
\hspace{1em} $\hat{\bm{\beta}}_{mm,block}$ & 0.0599 & 0.0283 &  &  & \tabularnewline
\hspace{1em} Post-ALasso$_{Sar,block}$ & 0.0076 & 0.0115 & 9.019 & 1.000 & 0.987\tabularnewline
\midrule 
\multicolumn{2}{l}{{\footnotesize{}$n=2,000$}} &  &  &  & \tabularnewline
\hspace{1em} Oracle 2sls & 0.0054 & 0.0080 & 9 & 1 & 1\tabularnewline
\hspace{1em} Naive 2sls & 0.3286 & 0.0107 & 0 & 0 & 0\tabularnewline
\hspace{1em} $\hat{\bm{\beta}}_{mm}$ & 0.0703 & 0.1667 &  &  & \tabularnewline
\hspace{1em} Post-ALasso$_{Sar}$ & 0.0071 & 0.1847 & 10.086 & 0.803 & 0.544\tabularnewline
\hspace{1em} $\hat{\bm{\beta}}_{mm,block}$ & 0.0411 & 0.0196 &  &  & \tabularnewline
\hspace{1em} Post-ALasso$_{Sar,block}$ & 0.0054 & 0.0084 & 9.020 & 1.000 & 0.987\tabularnewline
\bottomrule
\end{tabular}
\par\end{centering}
\medskip{}

{\footnotesize{}Notes: This table reports IV selection and estimation
results of the adaptive Lasso method for a design with a block structure
with no overlap. The reported statistics include median absolute error,
standard deviation, number of IVs selected as invalid, frequency with
which all invalid IVs have been selected as invalid, and frequency
with which oracle model has been selected. The simulations are based
on $1,000$ repetitions.}{\footnotesize\par}
\end{table}

\section{Application: The Effects of Educational Attainment and Cognitive
Ability on BMI}

\label{sec: application} We apply our IV selection method to a multivariable
Mendelian randomization (MVMR) study. We estimate the effects of educational
attainment and cognitive ability on Body Mass Index (BMI), as in \citet{SandersonetalIJE2019}.
Both educational attainment and cognitive ability have been found
to be negatively correlated with BMI \citep{SandersonetalIJE2019}.
However, as educational attainment and cognitive ability are highly
correlated, it is unclear to what extent each of them have a direct
effect on BMI. In this application, we account for both variables
in order to disentangle their direct effects. We use 74 SNPs as instruments
for educational attainment and 19 SNPs for cognitive ability, and
four SNPs overlap between the two sets of candidate instruments. These
SNPs have previously been identified in independent genome-wide association
studies, see \citet{okbay2016genome} for educational attainment,
and \citet{sniekers2017genome} for cognitive ability. We use data
on $86,150$ individuals of self-reported white European ancestry
from the UK Biobank. Educational attainment is measured as age participants
completed full-time education, imputed to 21 for individuals with
a degree. Cognitive ability is measured as a unitless fluid intelligence
score derived from tests completed during assessment and online followup.
We standardize the cognitive ability score to mean zero and variance
one. BMI is the ratio of weight in kilograms to height in metres squared,
both of which were measured for all individuals during assessment,
and we log-transform it due to skewness. Hence, we interpret our estimates
as the percentage change in BMI that is associated with a one unit
increase in the relevant explanatory variable. We also include additional
covariates that control for age at assessment, sex, and the first
10 genetic principal components, all of which are available from the
UK Biobank.\footnote{All genetic data passed quality control filters described here; https://research-information.bris.ac.uk/en/datasets/uk-biobank-genetic-data-mrc-ieu-quality-control-version-2 }
See \citet{SandersonetalIJE2019} for a detailed definition of the
variables and presentation of the data.

Table~\ref{tab:application} reports the results of our analysis.
Columns (1) and (2) show, respectively, the point estimates and their
standard errors. Column (3) is the number of instruments selected
as invalid, and column (4) shows the p-value of the Sargan test. The
first panel presents the estimates from a naive 2sls regression where
we treat all the candidate instruments as valid. The estimate for
the effect of education is statistically significant at the $1\%$
level, whereas the estimate for cognitive ability is significant at
the $5\%$ level, but not the $1\%$ level. However, these results
are from the naive 2sls regression, and they might be biased due to
the presence of invalid instruments. This is supported by the small
p-value of the Sargan test (\textless{} 2e-16). In practice, SNPs
can exhibit so-called pleiotropic effects, which would make them invalid
instruments. In our setting, pleiotropy would mean that some of the
SNPs, either for educational attainment or cognitive ability, have
direct effects on BMI.

\begin{table}[H]
\begin{centering}
\caption{\label{tab:application} The impacts of educational attainment and
cognitive ability on $\log(BMI)$}
\par\end{centering}
\medskip{}

\begin{centering}
\begin{tabular}{lrccc}
\toprule 
 & Estimate  & Std. error  & \# Invalid  & p-value, Sargan\tabularnewline
 & \multicolumn{1}{c}{(1)} & (2)  & (3)  & (4)2S\tabularnewline
\midrule 
\multicolumn{2}{l}{2sls} &  &  & \tabularnewline
\hspace{1em} Educational attainment  & -0.030  & 0.005  & 0  & \textless{} 2e-16\tabularnewline
\hspace{1em} Cognitive ability  & 0.032  & 0.013  &  & \tabularnewline
\midrule 
\addlinespace
Post-ALasso$_{Sar}$ &  &  &  & \tabularnewline
\hspace{1em} Educational attainment  & -0.026  & 0.005  & 12  & 0.009\tabularnewline
\hspace{1em} Cognitive ability  & 0.027  & 0.013  &  & \tabularnewline
\hspace{1em} $\hat{\beta}_{mm,edu}$  & -0.031  &  &  & \tabularnewline
\hspace{1em} $\hat{\beta}_{mm,cog}$  & 0.012  &  &  & \tabularnewline
\bottomrule
\end{tabular}
\par\end{centering}
\medskip{}

{\footnotesize{}Notes: The sample size is $n=86,150$. The number
of instruments for educational attainment is $k_{edu}=74$. The number
of instruments for cognitive ability is $k_{cog}=19$. There are four
instruments identified for both educational attainment and cognitive
ability, so the total number of putative instruments is $k_{z}=89$.
Median-of-medians estimator takes this block structure with overlap
into account.}{\footnotesize\par}
\end{table}

Instead of the naive 2sls, we now conduct IV selection using the adaptive
Lasso with the downward testing procedure, as described in Section~\ref{sec:downward},
and we obtain post-selection 2sls estimates. In the second panel of
Table~\ref{tab:application}, we report the results for the direct
effects of educational attainment and cognitive ability using our
method, and we show the estimates taking the block structure, with
overlap, into account. We also present the associated median-of-medians
estimates, denoted $\hat{\beta}_{mm,edu}$ and $\hat{\beta}_{mm,cog}$
for, respectively, educational attainment and cognitive ability. The
threshold p-value for the Sargan test is $0.1/\log(n)=0.0086$.

For educational attainment we find that $\hat{\beta}_{mm,edu}=-0.031$.
For cognitive ability, the estimate is $\hat{\beta}_{mm,cog}=0.012$.
We find that our method selects $12$ instruments as invalid. Nine
of these are for educational attainment and three are for cognitive
ability. As seen in Column (4), the p-value of the Sargan statistic
for the selected model is $0.009$. We find that the post-selection
2sls estimates are somewhat closer to zero compared to the estimates
for the naive 2sls. The post-selection estimate for educational attainment
is $-0.026$, while, for cognitive ability, it is $0.027$. The effect
of educational attainment on $\log$$\left(BMI\right)$ is still significant
at the $1\%$ level, while the effect of cognitive ability is again
not significant at the $1\%$ level. We therefore find limited evidence
of a direct effect of cognitive ability on BMI, as also indicated
by the median-of-medians estimator. For the results in Table~\ref{tab:application},
we assume conditional homoskedasticity. However, a robust version
of our method, i.e.\ using the two-step Hansen J-test (\citealp{Hansen1982})
and the post-selection two-step GMM estimator, produces virtually
identical results.

\section{Conclusions}

\label{sec: conclusion} We investigate the use of the adaptive Lasso
method for selecting valid instrumental variables from a set of candidate
instruments when some of the instruments may be invalid. While existing
work has focused on a single endogenous variable, our method contributes
to the literature by allowing for multiple endogenous exposure variables.
Under a modified majority rule, we show that the adaptive Lasso method
can achieve consistent selection and oracle estimation by using a
novel initial consistent estimator, the median-of-medians estimator.
In this work, we have considered the number of candidate instruments
to be fixed, but in some settings it may grow with the sample size
(or even at a faster rate), and future research will focus on extending
the method to handle such cases.

The standard errors we report in Table \ref{tab:application} are
the standard 2sls standard errors based on the oracle distribution
in (\ref{eq:thetaorlim}). As these do not take into account the selection
uncertainty, they are likely to underestimate the true variability
of the estimates, depending on the information in the data, see the
extensive analysis of the adaptive Lasso results for $k_{x}=1$ in
\citet{WindmeijeretalJASA2019}. In a recent paper, \citet{Guo2022}
proposed uniformly valid confidence intervals for the single exposure
selection methods of \citet{GuoetalJRSSB2018} and \citet{windmeijer2021confidence}.
Deriving uniformly valid confidence intervals for our proposed post-adaptive
Lasso estimator is an important topic for future research.

\section*{Acknowledgments}

We thank seminar participants at Bristol, Sheffield, Toronto and Toulouse
for helpful comments and suggestions. In particular, we thank Linbo
Wang for his suggestion to investigate the median-of-medians-of-medians
estimator for the $k_{x}=3$ case. Xiaoran Liang acknowledges support
from the the Economic and Social Research Council, ES/P000630/1. Eleanor
Sanderson works in a unit funded by the MRC (MRC\_UU\_00011/1). This
research has been conducted using the UK Biobank Resource under Application
Number 66074. The work was carried out using the computational facilities
of the Advanced Computing Research Centre, University of Bristol -
http://www.bristol.ac.uk/acrc/.

\section*{Appendix, Proofs}

\label{sec:app}

\subsection*{Proposition \ref{prop:bmj}}
\begin{proof}
The proof of Proposition \ref{prop:bmj}, with $k_{x}=2$, follows
the arguments of the proof of Theorem 1 in \citet{WindmeijeretalJASA2019}.
The estimands for the $k_{z}-1$ just-identified IV estimators in
the model specification (\ref{eq:jimod}) 
\[
\mathbf{y}=\mathbf{X}\boldsymbol{\beta}_{s}+\mathbf{Z}_{\left\{ -s\right\} }\bm{\alpha}_{\left\{ -s\right\} }+\mathbf{u}_{s},
\]
are given by $\boldsymbol{\beta}+\boldsymbol{\Pi}_{s}^{-1}\bm{\alpha}_{s}$,
with $\boldsymbol{\Pi}_{s}$ as defined in (\ref{eq:fss}). It then
follows, under Assumptions \ref{assexp}, \ref{clt} and \ref{ass:justrank}
that $\hat{\boldsymbol{\beta}}_{\ell}^{j}$ as defined in (\ref{eq:bmj})
is a consistent and normal estimator of $\boldsymbol{\beta}$ for
a valid instrument $j\in\mathcal{V}$ and when the other instrument
$\ell$ is also valid. For each element $\beta_{q}$ in $\boldsymbol{\beta}$,
$q=1,2$, we have for instrument $j$ a vector of $k_{z}-1$ estimators
$\hat{\boldsymbol{\beta}}_{q}^{j}$. Let $\boldsymbol{\delta}_{\ell}^{j}$
be the $k_{x}$-vector $\boldsymbol{\Pi}_{\left\{ j,\ell\right\} }^{-1}\bm{\alpha}_{\left\{ j,\ell\right\} }$,
with elements $\boldsymbol{\delta}_{\left\{ j,\ell\right\} ,q}^{j}$,
for $q=1,\ldots,k_{x}$. For each element we have the $\left(k_{z}-1\right)$-vector
$\boldsymbol{\delta}_{q}^{j}$. It follows that 
\[
\hat{\boldsymbol{\beta}}_{q}^{j}\stackrel{p}{\rightarrow}\beta_{q}\boldsymbol{\iota}_{k_{z}-1}+\boldsymbol{\delta}_{q}^{j},
\]
where $\boldsymbol{\iota}_{k_{z}-1}$ is a $\left(k_{z}-1\right)$-vector
of ones. For a valid instrument $j\in\mathcal{V}$ there are $k_{\mathcal{V}}-1$
sets with valid instruments only. By Assumption \ref{ass:jmajority}
$\left(k_{\mathcal{V}}-1\right)>\frac{k_{z}-1}{2}$, it follows that
the majority rule is satisfied and more than 50\% of the elements
of $\boldsymbol{\delta}_{q}^{j}$ are equal to zero. Using a continuity
theorem, it then follows that, for a valid instrument $j\in\mathcal{V}$,
\begin{equation}
\text{median}\left(\hat{\boldsymbol{\beta}}_{q}^{j}\right)\stackrel{p}{\rightarrow}\beta_{q}+\text{median}\left(\boldsymbol{\delta}_{q}^{j}\right)=\beta_{q},\label{eq:medplim}
\end{equation}
for $q=1,\ldots,k_{x}$, and hence the first result of Proposition
\ref{prop:bmj} therefore follows, $\hat{\boldsymbol{\beta}}_{m}^{j}\stackrel{p}{\rightarrow}\boldsymbol{\beta}$.

Under Assumptions \ref{assexp}, \ref{clt} and \ref{ass:justrank}
the limiting distribution of $\hat{\boldsymbol{\beta}}_{q}^{j}$,
for $q=1,\ldots,k_{x}$, is given by 
\[
\sqrt{n}\left(\hat{\boldsymbol{\beta}}_{q}^{j}-\left(\beta_{q}\boldsymbol{\iota}_{c}+\boldsymbol{\delta}_{q}^{j}\right)\right)\stackrel{d}{\rightarrow}N\left(\mathbf{0},\boldsymbol{\Sigma}_{\boldsymbol{\beta}_{q}^{j}}\right).
\]
For $\hat{\beta}_{m,q}^{j}=\text{median}\left(\hat{\boldsymbol{\beta}}_{q}^{j}\right)$
we have that 
\begin{align*}
\sqrt{n}\left(\hat{\beta}_{m,q}^{j}-\beta_{q}\right) & =\sqrt{n}\left(\text{median}\left(\hat{\boldsymbol{\beta}}_{q}^{j}\right)-\beta_{q}\right)\\
 & =\text{median}\left(\sqrt{n}\left(\hat{\boldsymbol{\beta}}_{q}^{j}-\beta_{q}\boldsymbol{\iota}_{k_{z}-1}\right)\right).
\end{align*}
For a valid instrument $j\in\mathcal{V}$, let $\boldsymbol{\delta}_{\mathcal{A},q}^{j}$
denote the $k_{z}-k_{\mathcal{V}}$ values of $\boldsymbol{\delta}_{q}^{j}$
for the sets that include invalid instruments and $\boldsymbol{\delta}_{\mathcal{V},q}^{j}=\mathbf{0}_{k_{\mathcal{V}}-1}$
the $k_{\mathcal{V}}-1$ values of $\boldsymbol{\delta}_{q}^{j}$
for the sets that only contain valid instruments. Partition $\boldsymbol{\delta}_{q}^{j}=\left(\left(\boldsymbol{\delta}_{\mathcal{A},q}^{j}\right)^{T}\,\,\mathbf{0}_{k_{\mathcal{V}}-1}^{T}\right)^{T}$
and equivalently $\hat{\boldsymbol{\beta}}_{q}^{j}=\left(\left(\hat{\boldsymbol{\beta}}_{\mathcal{A},q}^{j}\right)^{T}\,\,\left(\hat{\boldsymbol{\beta}}_{\mathcal{V},q}^{j}\right)^{T}\right)^{T}$.
Then 
\[
\sqrt{n}\left(\hat{\boldsymbol{\beta}}_{q}^{j}-\beta_{q}\boldsymbol{\iota}_{c}\right)=\left(\begin{array}{c}
\sqrt{n}\left(\hat{\boldsymbol{\beta}}_{\mathcal{A},q}^{j}-\left(\beta_{q}\boldsymbol{\iota}_{k_{z}-k_{\mathcal{V}}}+\boldsymbol{\delta}_{\mathcal{A},q}^{j}\right)\right)+\sqrt{n}\boldsymbol{\delta}_{\mathcal{A},q}^{j}\\
\sqrt{n}\left(\hat{\boldsymbol{\beta}}_{\mathcal{V},q}^{j}-\beta_{q}\boldsymbol{\iota}_{k_{\mathcal{V}}-1}\right)
\end{array}\right),
\]
and it follows that 
\begin{equation}
\sqrt{n}\left(\hat{\beta}_{m,q}^{j}-\beta_{q}\right)=\text{median}\left(\sqrt{n}\left(\hat{\boldsymbol{\beta}}_{q}^{j}-\beta_{q}\boldsymbol{\iota}_{k_{z}-1}\right)\right)\stackrel{d}{\rightarrow}H_{\left[r_{j}\right],k_{z}-1,q}^{j},\label{eq:asydist}
\end{equation}
for $q=1,\ldots,k_{x}$, and where, for $k_{z}-1$ odd, $H_{\left[r_{j}\right],k_{\mathcal{V}}-1,q}^{j}$
is the $r_{j}$th-order statistic of the limiting normal distribution
of $\sqrt{n}\left(\hat{\boldsymbol{\beta}}_{\mathcal{V},q}^{j}-\beta_{q}\boldsymbol{\iota}_{k_{\mathcal{V}}-1}\right)$,
where $r_{j}$ is determined by $k_{z}$, $k_{\mathcal{V}}$ and the
signs of the elements of $\boldsymbol{\delta}_{\mathcal{A},q}^{j}$.
For $k_{z}-1$ even, $H_{\left[r_{j}\right],k_{\mathcal{V}}-1,q}^{j}$
is defined as the average of either the $\left[r_{j}\right]$ and
$\left[r_{j}-1\right]$-order statistics, or the $\left[r_{j}\right]$
and $\left[r_{j}+1\right]$-order statistics, see \citet[p 1343]{WindmeijeretalJASA2019}.
From (\ref{eq:asydist}) it follows that $\hat{\beta}_{m,q}^{j}$
converges at the $\sqrt{n}$ rate. It has an asymptotic bias, but
$\sqrt{n}\left(\hat{\beta}_{m,q}^{j}-\beta_{q}\right)=O_{p}\left(1\right)$
for $q=1,2$, and so the second result of Proposition \ref{prop:bmj}
holds, $\sqrt{n}\left(\hat{\boldsymbol{\beta}}_{m}^{j}-\boldsymbol{\beta}\right)=O_{p}\left(1\right)$. 
\end{proof}

\subsection*{Proposition \ref{prop:mm}}
\begin{proof}
For $k_{x}=2$, and for $q=1,2$ we have the $k_{z}$ median estimators
$\hat{\beta}_{m,q}^{j}$, $j=1,\ldots,k_{z}$, of $\beta_{q}$. Denote
the $k_{z}$-vector of estimators by $\hat{\boldsymbol{\beta}}_{m,q}$.
Let $\hat{\boldsymbol{\beta}}_{m,q}^{\mathcal{V}}$ denote the $k_{\mathcal{V}}$-vector
$\left(\hat{\beta}_{m,q}^{j}\right)_{j\in\mathcal{V}}$ and $\hat{\boldsymbol{\beta}}_{m,q}^{\mathcal{A}}$
the $\left(k_{z}-k_{\mathcal{V}}\right)$-vector $\left(\hat{\beta}_{m,q}^{j}\right)_{j\in\mathcal{A}}$.
Partition $\hat{\boldsymbol{\beta}}_{m,q}=\left(\left(\hat{\boldsymbol{\beta}}_{m,q}^{\mathcal{A}}\right)^{T}\,\,\left(\hat{\boldsymbol{\beta}}_{m,q}^{\mathcal{V}}\right)^{T}\right)^{T}$.
Then under the assumptions and from the results of Proposition \ref{prop:bmj}
it follows that 
\[
\hat{\boldsymbol{\beta}}_{m,q}\stackrel{p}{\rightarrow}\left(\begin{array}{c}
\beta_{q}\boldsymbol{\iota}_{k_{z}-k_{\mathcal{V}}}+\boldsymbol{\gamma}_{q}\\
\beta_{q}\boldsymbol{\iota}_{k_{\mathcal{V}}}
\end{array}\right),
\]
where $\boldsymbol{\gamma}_{q}$ is the $\left(k_{z}-k_{\mathcal{V}}\right)$-vector
with elements $\text{median}\left(\boldsymbol{\delta}_{q}^{j}\right)_{j\in\mathcal{A}}$,
with $\boldsymbol{\delta}_{q}^{j}$ as defined in the proof of Proposition
\ref{prop:bmj}. Therefore, if the majority rule holds that $k_{\mathcal{V}}>\frac{k_{z}}{2}$,
it follows that 
\[
\beta_{mm,q}=\text{median}\left(\hat{\boldsymbol{\beta}}_{m,q}\right)\stackrel{p}{\rightarrow}\beta_{q},
\]
for $q=1,2$. From Assumption \ref{ass:jmajority} it follows $k_{\mathcal{V}}>\frac{k_{z}+1}{2}>\frac{k_{z}}{2}$,
and so it follows that the first result of Proposition \ref{prop:mm}
holds, $\boldsymbol{\beta}_{mm}\stackrel{p}{\rightarrow}\boldsymbol{\beta}$.

For the limiting distribution, we have, for $q=1,2$, 
\[
\sqrt{n}\left(\hat{\boldsymbol{\beta}}_{m,q}-\beta_{q}\boldsymbol{\iota}_{k_{z}}\right)=\left(\begin{array}{c}
\sqrt{n}\left(\hat{\boldsymbol{\beta}}_{m,q}^{\mathcal{A}}-\left(\beta_{q}\boldsymbol{\iota}_{k_{z}-k_{\mathcal{V}}}+\boldsymbol{\gamma}_{q}\right)\right)+\sqrt{n}\boldsymbol{\gamma}_{q}\\
\sqrt{n}\left(\hat{\boldsymbol{\beta}}_{m,q}^{\mathcal{V}}-\beta_{q}\boldsymbol{\iota}_{k_{\mathcal{V}}}\right)
\end{array}\right)
\]
and, as $k_{\mathcal{V}}>\frac{k_{z}}{2}$, 
\[
\sqrt{n}\left(\hat{\beta}_{mm,q}-\beta_{q}\right)=\text{median}\left(\sqrt{n}\left(\hat{\boldsymbol{\beta}}_{m,q}-\beta_{q}\boldsymbol{\iota}_{k_{z}}\right)\right)
\]
converges in distribution to the distribution of an order statistic
of the distribution of the order statistics $\left(H_{\left[r_{j}\right],k_{\mathcal{V}}-1,q}^{j}\right)_{j\in\mathcal{V}}$,
which is again $O_{p}\left(1\right)$. From this, the second result
of Proposition \ref{prop:mm} holds, $\sqrt{n}\left(\boldsymbol{\beta}_{mm}-\boldsymbol{\beta}\right)=O_{p}\left(1\right)$.
\end{proof}
\bibliographystyle{ecta}
\bibliography{selbib}

\end{document}